\documentclass[aps,prl,twocolumn,longbibliography,10pt,superscriptaddress]{revtex4-2}
\usepackage{graphicx}
\usepackage[english]{babel}
\usepackage{braket}
\makeatletter
\adddialect\l@en\l@english
\makeatother

\usepackage{amsmath,amssymb,braket}
\usepackage{tikz}
\usepackage[caption=false]{subfig}

\usepackage[percent]{overpic}
\usepackage{booktabs}
\usepackage{silence}
\usepackage[colorlinks=true,
            linkcolor=blue,
            citecolor=blue,
            urlcolor=blue]{hyperref}

\usepackage{xcolor}
\usepackage{amssymb}

\usepackage{amsthm}

\theoremstyle{plain}

\theoremstyle{definition}

\theoremstyle{remark}

\newtheorem*{assumption*}{Assumption}

\usepackage{microtype}
\usepackage{xspace}

\begin{document}

\title{Extensive mixed-state entanglement in kinetically constrained superradiance}
\author{Lucas Winter}
\affiliation{Faculty of Physics, University of Vienna, Boltzmanngasse 5, 1090 Vienna, Austria}
\author{Jan Kumlin}
\affiliation{Institute for Theoretical Physics, TU Wien, Wiedner Hauptstraße 8-10/136, 1040 Vienna, Austria}
\author{Thomas Pohl}
\affiliation{Institute for Theoretical Physics, TU Wien, Wiedner Hauptstraße 8-10/136, 1040 Vienna, Austria}
\author{Andreas Nunnenkamp}
\affiliation{Faculty of Physics, University of Vienna, Boltzmanngasse 5, 1090 Vienna, Austria}
\date{\today}

\begin{abstract}
Dicke superradiance by an ensemble of quantum emitters produces a collective burst of radiation, but no entanglement in the mixed state of the emitters.
We show that adding a local kinetic constraint between the emitters generates extensive mixed-state entanglement, while otherwise preserving all key features of Dicke superradiance. Specifically, for any local Boolean constraint, we analytically derive a lower bound for the emission rate which implies a peak intensity $\propto N^2$ and a peak time $\propto (\log N)/N$ with the number of spins $N$. This effect enables the superradiantly accelerated preparation of entangled dark states. Hereby, Hilbert-space fragmentation of the Dicke ladder leads to an exponentially branching decay tree that generates a hierarchy of dark states. Importantly, these disconnected manifolds include exponentially many long-range entangled singlet dark states. The explored kinetic constraints and superradiant dynamics can be realized in neutral-atom arrays coupled to an optical cavity, and we suggest a simple and accessible witness to detect the predicted mixed-state entanglement in such experiments. Moreover, we show that entanglement generation is robust against atomic decay and collective dephasing, and should be observable under recently reported experimental conditions. Our results, thereby, offer a general framework and experimentally viable approach for the dissipative engineering of entangled dark states enhanced by superradiance.
\end{abstract}

\maketitle

\emph{Introduction}---Superradiance is a hallmark of collectivity in quantum optics. Global coherence and the synchronization of transition dipoles among emitters produce a burst of radiation~\cite{Dicke1954, RehlerEberly1971, SkribanowitzEtAl1973,GrossHaroche1982,TavisCummings1968,MassonAsenjoGarcia2022, SierraMassonAsenjoGarcia2022} that becomes progressively stronger and shorter as the number of particles increases.
Signatures of superradiant dynamics have been observed in a wide range of physical systems, from superradiant lasers \cite{Bohnet2012}, and bosonic \cite{Baumann2010} and fermionic \cite{Zhang2021} quantum gases in optical cavities, to solid-state systems \cite{Mlynek2014}. 
The long-standing question whether Dicke superradiance produces quantum  entanglement was only answered very recently: although individual quantum trajectories can be highly entangled, the unconditional density matrix remains separable at all times~\cite{Rosario2025CSS,Bassler2025NoEnt,ZhangMalzRabl2025UnravelingSR}. Conventional superradiance cannot generate mixed-state entanglement on a fast time scale. 


Additional particle interactions provide a direct means of generating quantum correlations. The combination of collective light-matter coupling and short-range particle interactions is currently attracting substantial interest and has motivated recent studies of the so-called Dicke-Ising model. These studies have revealed rich physics, ranging from interesting ground-state phases~\cite{Zhang2013,DongEtAl2026DickeIsing,LangheldHormannSchmidt2025DickeIsing, Koziol2025Melting} to exotic many-body scars~\cite{HosseinabadiEtAl2025PXP2} and string dynamics~\cite{PuelMacri2024Mesons,BacciconiEtAl2025LocalNonlocal}, driven by the interplay of light-induced long-range coupling and short-range interactions between emitters. Yet, the consequences of this competition for the dissipative dynamics and in particular for superradiance in interacting emitter ensembles have remained largely unexplored.  

\begin{figure}[!t]
  \centering
  \begin{tikzpicture}
    \node[anchor=south west, inner sep=0] (setupfig) at (0,0)
      {\includegraphics[width=\columnwidth, trim=-3in 0in 3in 0in,clip]{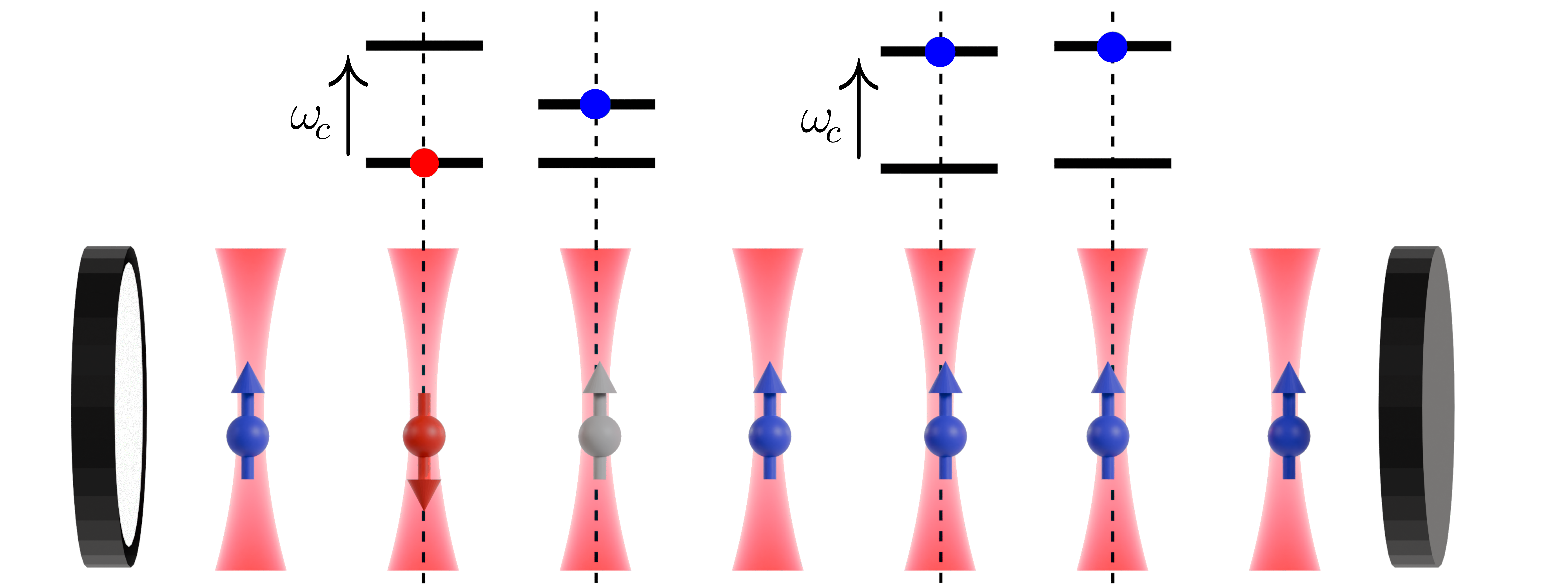}};
    \node[anchor=north west, fill=white, inner sep=1pt] at ([xshift=4pt,yshift=-4pt]setupfig.north west) {\normalsize (a)};
  \end{tikzpicture}
  \includegraphics[width=\columnwidth]{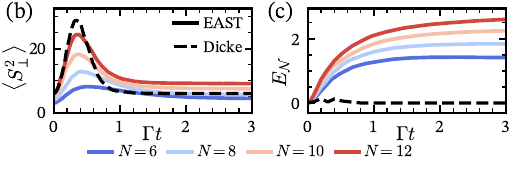}
  \par \vspace{-0.6em}
 \caption{\textbf{Rydberg-blockaded superradiance.}
(a)~We consider $N$ atoms coupled to a cavity mode of frequency $\omega_c$.
A Rydberg (anti-)blockade locally gates the atom-cavity transition \eqref{eq:P};
for the EAST constraint, $P_j=n_{j-1}$, atom $j$ couples only if its left neighbor is excited.
(b,c)~Dynamics of the Rydberg Tavis--Cummings model \eqref{eq:H} (solid) and the Dicke model (black dashed) for different system sizes.
Panel (b) shows the transverse spin coherence $\langle S_\perp^2\rangle=\langle S_x^2+S_y^2\rangle$; panel (c) shows mixed-state entanglement quantified by logarithmic negativity $E_{\mathcal N}$ \eqref{eq:logneg}.
Quantum trajectory simulations for $\kappa/g=40$, $\Delta=0$, and $N_{\text{traj}}=600$.}
\label{fig:setup}
\end{figure}

In this letter, we show that kinetic constraints convert superradiance from a non-entangling collective decay into a fast dissipative route toward highly entangled dark states. 
Such kinetic constraints are generated by strong short-range interactions \cite{JackleEisinger1991,LesanovskyGarrahan2013,ValenciaTortoraEtAl2024} and break the strict permutation symmetry underlying Dicke superradiance. Here, we show analytically that -- despite breaking permutation symmetry -- \emph{any local Boolean constraint in arbitrary dimensions} does not diminish superradiance, but retains the superradiant burst with its characteristic $\propto N^2$ intensity scaling and fast time scale $\propto (\log N)/N$ that decreases with the number of emitters [Fig.~\ref{fig:setup}(b)]. At the same time, we demonstrate numerically that entanglement emerges during the constrained decay dynamics and grows linearly with the number of emitters [Fig.~\ref{fig:setup}(c)]. This is traced back to Hilbert-space fragmentation \cite{SalaEtAl2020,KhemaniHermeleNandkishore2020,MoudgalyaBernevigRegnault2022,YangEtAl2020StrictConfinement,BrighiLjubotinaSerbyn2023QuantumEast,LiSalaPollmann2023Open,BrighiLjubotina2024EastWest,PaszkoEtAl2025,GarciaGarciaEtAl2026LindbladScars} and the emergence of disconnected dark manifolds containing entangled $N$-body states. Remarkably, these findings imply that \emph{kinetically constrained superradiance generates extensive entanglement on a fast superradiantly accelerated timescale}. We present numerical calculations for three different kinetic constraints that can be realized in experiments with optical-tweezer arrays of Rydberg atoms~\cite{LabuhnEtAl2016,BernienEtAl2017,BrowaeysLahaye2020,ValenciaTortoraEtAl2024}, for which the coupling to an optical cavity has just been demonstrated experimentally \cite{SantisRydbergCavity2026}. Importantly, we find that the discovered effect is robust to inevitable imperfections, such as laser phase-noise and spontaneous atomic decay, yielding near-optimal entanglement with experimentally achievable cooperativities of the atom-cavity setup. Together, our results suggest a new concept and viable route for the dissipative and superradiantly enhanced generation of extensive entanglement in mixed quantum states of large systems. 

\begin{figure}[!t]
  \centering
  \includegraphics[width=\columnwidth,trim=0.0in -0.0in 0.05in 0.0in,clip]{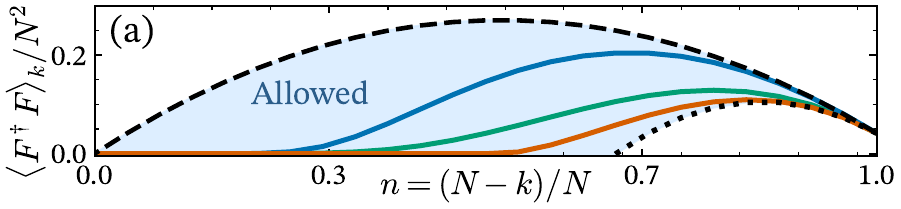}
  \includegraphics[width=\columnwidth,trim=0.0in 0.0in -0.0in 0.0in,clip]{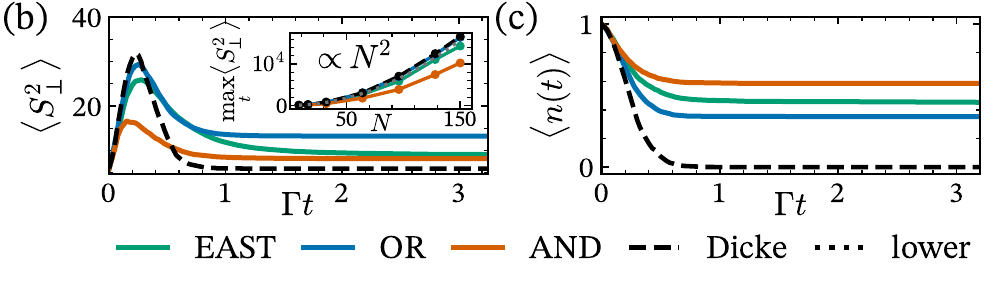}
  \par \vspace{-0.6em}
 \caption{\textbf{Emission rate and constrained superradiant dynamics.}
(a)~Emission rate $\langle F^\dagger F\rangle_k/N^2$ \eqref{eq:FdagF} as a function of $n = 1-k/N$ for $N=24$.
The shaded region is the allowed window for local Boolean constraints including AND (orange), EAST (green) and OR (blue) with an upper bound given by the Dicke model and a lower bound given by Eq.~\eqref{eq:lowerBound}.
(b)~Exact dynamics of $\langle S_\perp^2\rangle=\langle S_x^2+S_y^2\rangle$ according to Eq.~\eqref{eq:masterEq}. All constrained models retain a superradiant burst.
Inset: Peak heights scale as $\max_t\langle S_\perp^2\rangle\propto N^2$ according to DTWA simulations (see SM \cite{SI}).
(c)~Mean excitation density $\braket{n(t)} = \braket{S^z}/N + 1/2$. The Dicke model decays to the ground state, while constraints arrest the decay at finite density.
Numerical simulations in (b,c) use quantum trajectory method \cite {Castin, Knight}, with $N=12$, $\Delta=0$, $\kappa/g=30$, and $N_{\rm traj}=200$ trajectories.}
\label{fig:old_setup_panels}
\end{figure}

\begin{figure*}[t]
  \centering
  \subfloat[]{\begin{overpic}[width=0.42\textwidth]{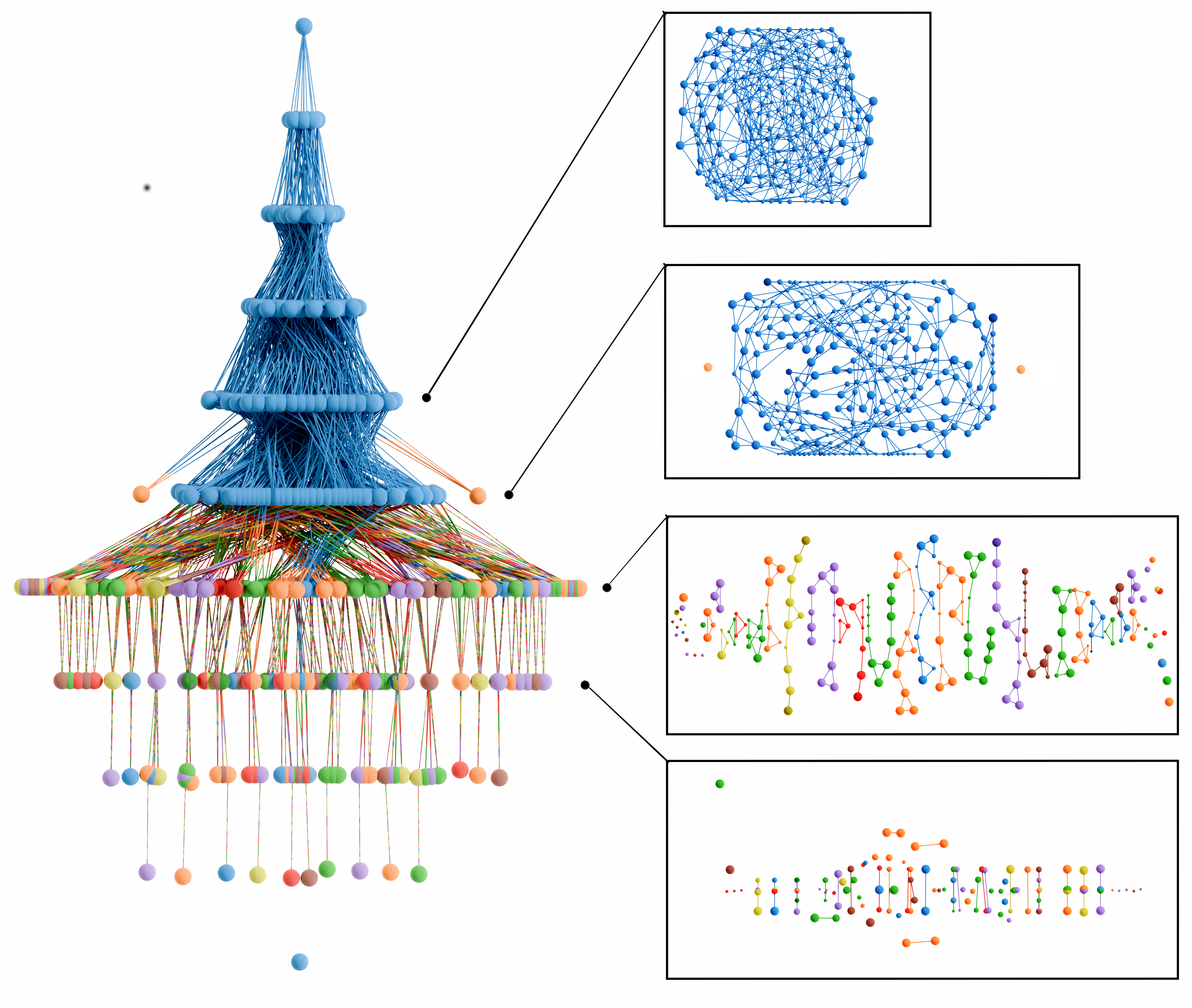}
    \put(4,84){\normalsize (a)}
  \end{overpic}}\hfill
  \subfloat[]{\begin{minipage}[t]{0.54\textwidth}
    \centering
    \includegraphics[width=\linewidth]{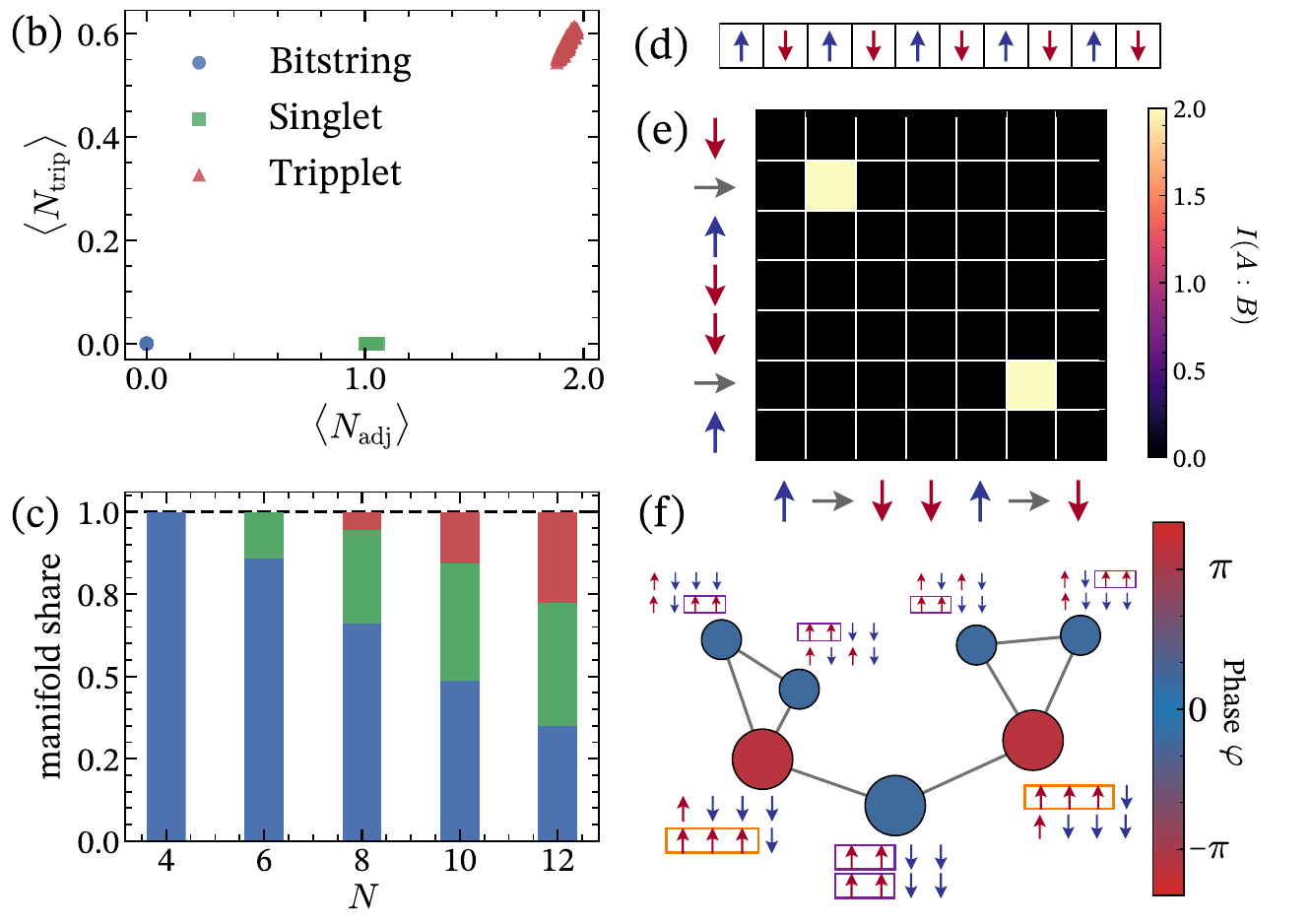}
  \end{minipage}}
 \caption{\textbf{Hilbert-space fragmentation and dark-manifold structure.} 
(a)~Numerically constructed Lindblad connectivity graph for $N=10$. Nodes represent spin configurations in the computational basis, grouped into planes by total magnetization $S^z$. The non-hermitian Hamiltonian $H_{\text{nh}} = \left(\chi - i\Gamma/2\right) F^\dagger F$ connects states in-plane (insets), while the jump operator $L_{\text{eff}}$ \eqref{eq:Heff} connects states vertically.
At small $S^z$, the graph fragments into many disconnected components (shown in different colors).
(b)~Dark state classification obtained from exact diagonalization of $F^\dagger F$ and identification of its kernel for $N=10$. Local adjacent-pair $N_{\rm adj}$ and triple $N_{\rm tri}$ counts categorize the dark state manifold into bitstrings (blue), singlet-decorated states (green), and triple states (red).
(c)~Dark-manifold composition. As system size $N$ increases, the bitstring fraction decreases while entangled singlet and triple manifold weight increases. 
(d)~Representative bitstring dark state from the numerically obtained kernel. 
(e)~Mutual-information matrix $I_{ij}=S[\mathrm{Tr}_{\overline{i}}\rho]+S[\mathrm{Tr}_{\overline{j}}\rho]-S[\mathrm{Tr}_{\overline{ij}}\rho]$, where $S$ is the von Neumann entropy, for a representative singlet dark state from the same kernel, revealing long-range entanglement.
(f)~Representative triple dark state from the kernel. Destructive interference between configurations with three consecutive excitations stabilizes a dark superposition (node size indicates amplitude, color encodes phase $\varphi$).}
  \label{fig:two_panel}
\end{figure*}

\emph{Model}---We consider a chain of $N$ Rydberg atoms coupled to a single cavity mode,
which we describe with an effective Rydberg Tavis--Cummings model 
\begin{align}
    H = \hbar \Delta a^\dagger a +  \hbar g (F a^\dagger + F^\dagger a). \label{eq:H}
\end{align}
Here, $a$ ($a^\dagger$) annihilates (creates) a cavity photon, $g$ is the atom-cavity coupling, and $\Delta$ the detuning between cavity frequency and the atomic transition frequency.
 $F$ is the constrained collective spin lowering operator,
\begin{align}
    F = \sum_j P_j \sigma_j^- . \label{eq:P}
\end{align}
Here, $\sigma_j^-$ is the  spin-$1/2$ lowering operator for site $j$ and the projector $P_j$ enforces the kinetic constraint induced by Rydberg (anti-)blockade: an atom only couples to the cavity depending on the state of neighboring atoms. Specifically, $P_j$ can be any Boolean constraint of range $w$ built from the local occupations $n_{j-w},\ldots,n_{j+w}$ around site $j$ with $n_j=(1+\sigma_j^z)/2$.
As an example, Figure~\ref{fig:setup}(a) illustrates the EAST model, $P_j=n_{j-1}$, where site $j$ couples only to the cavity if the left neighbor $j-1$ is excited. We also consider the AND model, $P_j=n_{j-1}n_{j+1}$ and the OR model, $P_j=n_{j-1}+n_{j+1}-n_{j-1}n_{j+1}$.
The unconstrained Dicke model is recovered by setting $P_j=1$. We use periodic boundary conditions throughout this work.

To account for open system dynamics, the density matrix $\rho$ evolves according to the master equation
\begin{align}
  \dot\rho = -\frac{i}{\hbar}[H,\rho] + \kappa \mathcal{D}[a]\rho, \label{eq:masterEq}
\end{align}
where the standard dissipator $\mathcal{D}[L]\rho = L\rho L^{\dagger} - \frac{1}{2}\{L^{\dagger}L,\rho\}$ describes single-photon loss at rate $\kappa$.

\emph{Early-time superradiance}---A first central question is whether the superradiant intensity scaling persists despite the blockade. We address this analytically in the bad-cavity regime $\Delta^2 + \kappa^2\gg g^2 N$.
Adiabatic elimination of the cavity (see Supplemental Material~\cite{SI}) yields an effective spin-only description
\begin{align}
  H_{\mathrm{eff}} &= \hbar \chi\, F^{\dagger}F,
  \qquad
  L_{\mathrm{eff}} = \sqrt{\Gamma}\, F, \label{eq:Heff}
\end{align}
with \(\chi = -\frac{4g^{2}\Delta}{4\Delta^{2}+\kappa^{2}}\) and \(\Gamma = \frac{4g^{2}\kappa}{4\Delta^{2}+\kappa^{2}}\).
The Hamiltonian term \(F^{\dagger}F\) generates constrained long-range flip-flop interactions, and \(L_{\mathrm{eff}}\) describes constrained collective loss.

We derive the equation of motion for the mean excitation density $\braket{n}=\sum_j \braket{n_j}/N$, where $n_j=(1+\sigma_j^z)/2$ are the occupation operators. In the resonant case $\Delta=0$, the dynamics is purely dissipative with $\chi=0$, giving
\begin{align}
    \partial_t \langle n \rangle = -\frac{\Gamma}{N}\langle F^\dagger F\rangle \, .
\end{align}
We note that this equation has the same form as in standard Dicke superradiance where the collective lowering operator $S^-$ replaces the constrained operator $F$.
Importantly, this form is valid for any local Boolean constraint $P(n_{j-w}, \dots, n_{j+w})$, as the constraint commutes with $n$. 

We now show that the asymptotic scaling of the emission rate $\braket{F^\dagger F}$ remains quadratic for any local Boolean constraint, as in Dicke superradiance. We use a click-limit approximation valid at early times by neglecting the non-Hermitian evolution between emissions. The state after $k$ jumps is then approximated by $|\psi_k\rangle = F^k |\!\uparrow \uparrow\dots \rangle$, and the emission rate is determined by the ratio of consecutive norms,
\begin{align}
    \langle F^\dagger F\rangle_k
    = \langle \psi_k | F^\dagger F|\psi_k\rangle
    = \frac{\langle \psi_{k+1} | \psi_{k+1} \rangle}{\langle \psi_k | \psi_k \rangle} \, . \label{eq:FdagF}
\end{align}

The precise form of $\braket{F^\dagger F}_k$ depends on the Boolean constraint and is suppressed more strongly by more restrictive $P_j$; see Fig.~\ref{fig:old_setup_panels}(a). At early times, however, this dependence is subleading:
only a dilute set of sites has decayed from the initially
fully excited state, and the dynamics has not yet resolved
the local constraint. In the Supplemental Material~\cite{SI}, we make this statement quantitative by proving the lower bound
\begin{align}
  \begin{split}
\braket{F^\dagger F}_k &\geq (k+1)\bigl[N-(2w+1)k\bigr]\propto \Theta(N^2) \ .\label{eq:lowerBound}
  \end{split}
\end{align}
This bound shows that the peak intensity remains quadratic, $\braket{F^\dagger F}_{\rm peak}\propto N^2$. It also preserves the superradiant acceleration of the emission, with $t_{\rm peak}\propto(\log N)/N$ (see Supplemental Material~\cite{SI}). The robustness of both the peak intensity and the peak time is our first main result and applies broadly to local Boolean constraints in arbitrary spatial dimensions. 

We now turn to numerically exact quantum-trajectory simulations of the full Lindblad dynamics \eqref{eq:masterEq}. In Figure~\ref{fig:old_setup_panels}(b) we show the transverse spin coherence, $\langle S_\perp^2 \rangle = \langle S_x^2 + S_y^2 \rangle$, for the unconstrained Dicke model (dashed) and the AND, EAST, and OR models (solid) in Fig.~\ref{fig:old_setup_panels}(b).
The constraint suppresses the peak height relative to the Dicke case, but the numerically extracted peak is consistent with the analytical quadratic scaling $\propto N^2$ derived above.

\emph{Late-time dark states}---Figure~\ref{fig:old_setup_panels}(c) shows that constrained models qualitatively change the stationary state. We plot the mean excitation density, $\braket{n}=\sum_j\braket{n_j}/N$, for the Dicke, AND, EAST, and OR models. At early times, the curves nearly coincide: few atoms have
decayed, and most sites still couple to the cavity. At late times, the Dicke model relaxes to the ground state, whereas constrained models retain a finite residual excitation density. This behavior is consistent with the analysis of $\braket{F^\dagger F}_k$ [Fig.~\ref{fig:old_setup_panels}(a)]: once the excitation density is no longer close to one, the constraint can strongly suppress emission. More restrictive constraints, such as AND, therefore arrest the decay earlier, while weaker constraints, such as OR, allow the system to decay further.

Hilbert-space fragmentation accounts for the late-time arrest of relaxation. Here we focus on the EAST constraint, but qualitatively similar results also apply to the AND and OR models. Figure~\ref{fig:two_panel}(a) shows the numerically constructed connectivity graph of the effective Lindbladian in the computational basis: each node is a bitstring $\ket{\{s_j\}}$, and nodes are grouped into planes of fixed total magnetization $S^z$. Since the non-Hermitian Hamiltonian $H_{\text{nh}} = \left(\chi - i\Gamma/2\right) F^\dagger F$ conserves particle number, it generates connections between basis states only within a given $S^z$ plane.
For smaller $S^z$, the resulting in-plane adjacency graph splits into exponentially many disconnected components, i.e., classical Hilbert-space fragmentation \cite{SalaEtAl2020,KhemaniHermeleNandkishore2020,MoudgalyaBernevigRegnault2022}.
The dissipator $L_{\mathrm{eff}}\propto F$ introduces interplane (vertical) connections between basis states by lowering $S^z$, thereby coupling fragments in sector $S^z$ to fragments in sector $S^z-1$. Unlike the Dicke ladder, where each excited state decays to a unique lower state \cite{Dicke1954,GrossHaroche1982}, fragmentation yields an exponentially branching, tree-like decay structure.
This gives rise to a hierarchy of dark states.

The simplest dark states are blockade-enabled. For $S^z\le N/2$, this tree terminates in absorbing leaves: simple product states with no adjacent excitations,
\begin{align}
\ket{\psi_{\mathrm{bit}}}=\ket{\{s_j|\  s_j s_{j+1}=0\}}, \label{eq:darkbitstring}
\end{align}
which are annihilated by both the jump operator $F=\sum_j n_{j-1}\sigma_j^-$ and the non-Hermitian Hamiltonian $H_{\text{nh}}\propto F^\dagger F$, i.e., $F\ket{\psi_{\mathrm{bit}}}=0$; see Fig.~\ref{fig:two_panel}(d) for an example.

Collective dissipation generates a second class of dark states beyond the bitstring manifold through destructive interference~\cite{Dicke1954,PlenioHuelgaBeigeKnight1999,KrausEtAl2008}. Starting from an EAST bitstring dark state \(\ket{\psi_{\rm bit}}\), one can dress an allowed pair of sites with a local antisymmetric excitation,
\begin{equation}
p_{ij}\ket{\psi_{\rm bit}}
= n_{i-1}n_{j-1}\big(\sigma_i^{+}-\sigma_j^{+}\big)\ket{\psi_{\rm bit}}, 
\label{eq:darksinglet}
\end{equation}
and obtain another dark state. The constrained collective jump operator \(F=\sum_k n_{k-1}\sigma_k^{-}\) couples only to the symmetric bright combination, so the antisymmetric pair is annihilated by the decay. This dressing can be repeated multiple times provided the generators have non-overlapping support $[p_{ij}, p_{kl}]=0$. This generates a family of singlet-decorated dark states from a single bitstring root. The dark manifold therefore contains entangled singlet tilings driven by destructive interference; see Fig.~\ref{fig:two_panel}(e) for an example.

The constraint also stabilizes a third class of entangled dark states with support on configurations containing at least three consecutive excitations. To distinguish the different dark-state families, we characterize the numerically obtained dark manifold $\ker(L_{\mathrm{eff}}^{\dagger}L_{\mathrm{eff}})$ using the adjacent pair and triple number operators
\begin{align}
  N_{\mathrm{adj}} &= \sum_j n_j n_{j+1},  \label{eq:Nadj}\\
  N_{\mathrm{tri}} &= \sum_j n_j n_{j+1} n_{j+2}.
\end{align}
Figure~\ref{fig:two_panel}(b) separates the dark manifold into three clearly distinct clusters. Bitstring dark states (blue) satisfy $\langle N_{\mathrm{adj}}\rangle=\langle N_{\mathrm{tri}}\rangle=0$, consistent with their product-state character. Singlet-decorated states (green) have a finite number of adjacent pairs, $\langle N_{\mathrm{adj}}\rangle>0$, while not having adjacent triple excitations $\langle N_{\mathrm{tri}}\rangle\simeq 0$. Figure~\ref{fig:two_panel}(c) shows that, as the system size increases, a growing fraction of the dark manifold consists of entangled states beyond the simple bitstring sector.

The remaining states (red), characterized by $\langle N_{\mathrm{tri}}\rangle>0$, lie beyond both the bitstring and singlet constructions. They arise from the interplay of blockade and destructive interference: configurations with three consecutive excitations are coherently superposed with nearby pair configurations so that all allowed decay amplitudes cancel. Figure~\ref{fig:two_panel} (f) shows one representative dark state living in a smaller Hilbert-space fragment. In the Supplemental Material, we give an analytical construction of the entire hierarchy of dark states.

\begin{figure}
  \centering
  \includegraphics[width=\columnwidth]{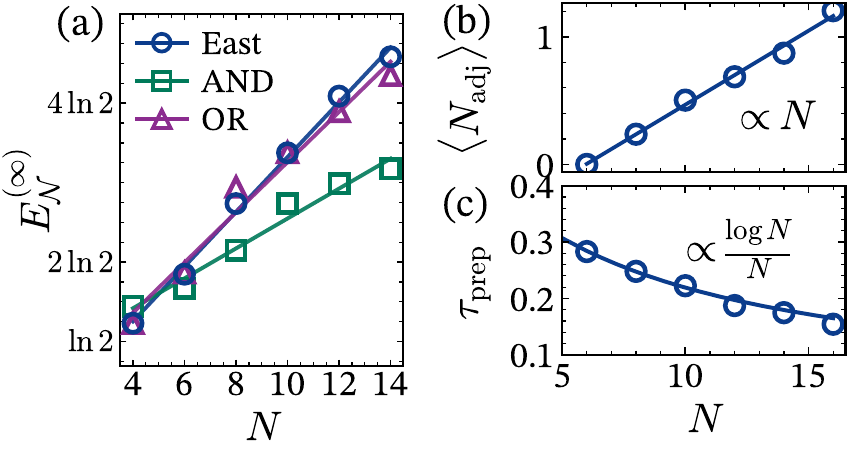}
  \caption{\textbf{Superradiantly accelerated generation of extensive mixed-state entanglement.}
(a) Stationary logarithmic negativity $E_{\mathcal N}^{(\infty)}$ \eqref{eq:logneg} across a half-chain bipartition, obtained from density matrices reconstructed from quantum-trajectory simulations of the effective spin model \eqref{eq:Heff}. The scaling is approximately linear in $N$ for the EAST (blue), OR (purple), and AND (green) constraints, showing extensive mixed-state entanglement.
(b) Adjacent pair count $\langle N_{\mathrm{adj}}\rangle \equiv\sum_j \langle n_j n_{j+1}\rangle$ in the stationary state, extracted from the same quantum-trajectory simulations, grows linearly with system size, $\langle N_{\mathrm{adj}}\rangle \propto N$, providing a direct experimental probe of mixed-state entanglement.
(c) Preparation time $\tau_{\mathrm{prep}}$, defined numerically from the approach to stationarity, versus system size. The numerically extracted data follow $\tau_{\mathrm{prep}}\propto (\log N)/N$, showing that collective emission accelerates the preparation of correlated dark states.
Quantum-trajectory simulations for $\kappa/g=40$, $\Delta=0$, and $N_{\text{traj}}=600$.
}
  \label{fig:negativity}
  
\end{figure}

\emph{Superradiantly accelerated extensive entanglement}---A central question is whether the stationary state is more than a trivial mixture of bitstring dark states and instead retains coherent quantum correlations. We perform exact quantum-trajectory simulations of the effective spin dynamics, reconstruct the density matrix from sufficiently many samples, and compute the logarithmic negativity \(E_{\mathcal N}\) across a bipartition \(A|B\) (here \(A\) denotes the left half of the chain). Note that the von Neumann entropy of the density matrix or the trajectory entanglement entropy does not by itself measure mixed-state entanglement, since it can be nonzero even for a separable density matrix~\cite{NielsenKempe2001}, as is the case in the standard Dicke model~\cite{Rosario2025CSS,Bassler2025NoEnt,ZhangMalzRabl2025UnravelingSR}. 
 For a density matrix \(\rho_{AB}\), the logarithmic negativity is defined as
\begin{equation}
E_{\mathcal N}(\rho_{AB}) \equiv \log \big\|\rho_{AB}^{T_B}\big\|_{1}, \label{eq:logneg}
\end{equation}
where \(\rho_{AB}^{T_B}\) is the partial transpose with respect to subsystem \(B\), \(\|\cdot\|_1 \equiv \mathrm{Tr}\sqrt{X^\dagger X}\) is the trace norm, and we use the natural logarithm. \(E_{\mathcal N}\) vanishes for separable states, and \(E_{\mathcal N}>0\) certifies mixed-state entanglement \cite{VidalWerner2002, Plenio2005LogNeg}.

The emission dynamics prepare a robust, extensively entangled stationary state according to finite-size quantum-trajectory simulations. Figure~\ref{fig:negativity}(a) shows that the stationary logarithmic negativity grows approximately linearly with system size, \(E_{\mathcal N}^{(\infty)}\propto N\), implying extensive bipartite entanglement across the cut for accessible system sizes. This rules out a stationary state that is merely a classical mixture of product (bitstring) dark states with \(E_{\mathcal N}=0\). Instead, the stationary density matrix must retain coherent off-diagonal structure within the dark manifold. By contrast, in conventional (permutation-invariant) Dicke superradiant decay, the unconditional density matrix remains separable, admitting a positive decomposition into coherent spin states, and hence has vanishing logarithmic negativity at all times \cite{Rosario2025CSS, Bassler2025NoEnt, ZhangMalzRabl2025UnravelingSR}. The logarithmic negativity scaling is generic beyond the EAST model and also applies to the OR and AND models. 
This highlights the special role of Rydberg interactions, which break permutation symmetry and generate genuine mixed-state entanglement.

A directly experimentally accessible witness of stationary-state entanglement within the dark manifold is the adjacent pair count $\langle N_{\rm adj}\rangle$ [Eq.~\eqref{eq:Nadj}]. The intuition is simple: any separable stationary dark state is a statistical mixture of bitstring dark states, and every bitstring dark state satisfies $n_j n_{j+1}=0$ on every bond [Eq.~\eqref{eq:darkbitstring}]. Such mixtures therefore necessarily give $\langle N_{\rm adj}\rangle=0$. By contrast, nonzero $\langle N_{\rm adj}\rangle$ requires overlap with entangled dark states such as the singlet and triple dark states and thus signals entanglement. In the Supplemental Material, we prove this statement rigorously: any separable stationary state obeying $\mathrm{Tr}(F^\dagger F\,\rho_{\rm stat})=0$ must also satisfy $\mathrm{Tr}(\rho_{\rm stat} N_{\rm adj})=0$. Figure~\ref{fig:negativity}(b) shows that $\langle N_{\rm adj}\rangle$ grows linearly with system size in the stationary state, showing entanglement for system sizes beyond where logarithmic negativity can be computed and providing an experimentally measurable probe \cite{LabuhnEtAl2016,BernienEtAl2017,BrowaeysLahaye2020}.

Early-time superradiant dynamics accelerate the preparation of the entangled stationary state. Figure~\ref{fig:negativity}(c) shows the preparation time, defined numerically as the time required for the adjacent excitation count $\braket{N_{\mathrm{adj}}}$ to reach 70 percent of its stationary value, as a function of system size $N$. A fit to the numerically extracted data reveals $(\log N)/N$ scaling, consistent with our analytical prediction for the superradiant burst time. This favorable scaling avoids the notoriously slow subradiant relaxation that usually prevents dark-state preparation.

\textit{Conclusion}---We have shown that local kinetic constraints in quantum-emitter ensembles retain Dicke superradiance while generating extensive mixed-state entanglement on a rapid superradiantly accelerated time scale.
This effect results from the fragmentation of the Dicke ladder into a branching decay tree with exponentially many absorbing leaves (Fig.~\ref{fig:two_panel}). The studied kinetic constraints can be realized via the Rydberg blockade in neutral-atom tweezer arrays~\cite{LabuhnEtAl2016,BernienEtAl2017,BrowaeysLahaye2020,ValenciaTortoraEtAl2024} for which collective cavity coupling has just been demonstrated experimentally~\cite{SantisRydbergCavity2026}. Our results provide a general framework for engineering large entangled dark states and suggest a feasible experimental implementation based on recently demonstrated technologies, thereby opening new directions for future research. For example, the possibility of tailoring specific dark states by programming the atom–cavity coupling beyond the symmetric Dicke model, or of controlling the resulting nonclassical properties of the emitted light, may be promising avenues for further studies.

\textit{Note added}---While finalizing this manuscript, a related preprint on superradiance and kinetic constraints appeared on the arXiv~\cite{DosPrazeresHosseinabadiMarino2026KCSuperradiance}. There, the authors consider average trajectory entanglement, while our work reports on mixed-state entanglement of the unconditional density matrix. 

\begin{acknowledgments}
\textit{Acknowledgments}---We thank P.~Borchia, J.~Koziol, D.~Malz, Y.~Minoguchi, and C.~Wanjura for fruitful discussions.
We thank P.~Brighi for detailed comments on the draft.
This research was funded in whole or in part by the Austrian Science Fund (FWF) [10.55776/COE1, 10.55776/F101200] and the European Union (NextGenerationEU), the European Union’s Horizon Europe research and innovation program under the Marie Sklodowska-Curie Grant Agreement No.~101106552 (QuLowD), and the European Research Council through the ERC Synergy Grant SuperWave (Grant No.~101071882).
\end{acknowledgments}

\bibliography{bib}

\onecolumngrid
\section*{End Matter}
\twocolumngrid

\begin{figure}[t]
  \centering
  \includegraphics[width=\columnwidth]{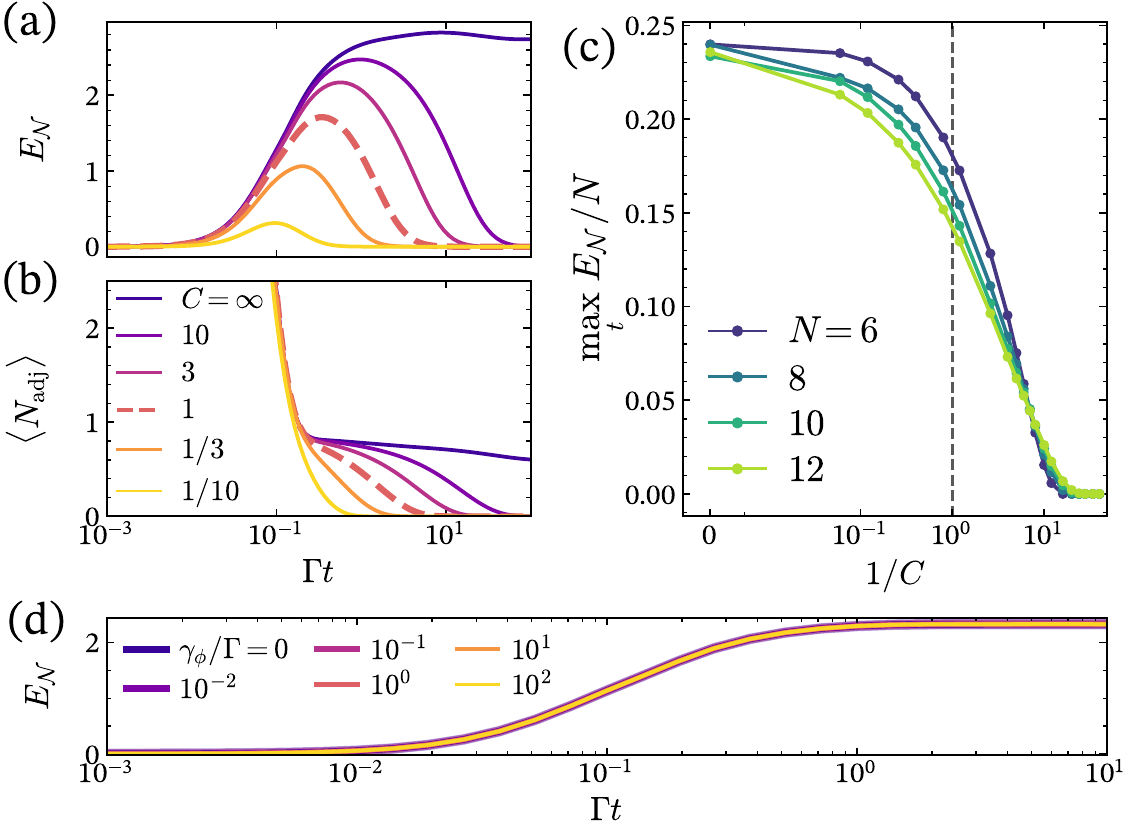}
  \caption{\textbf{Robustness to single-site loss and common-mode dephasing.}
  (a,b) Time evolution of the half-chain logarithmic negativity $E_{\mathcal N}$ and adjacent pair count $\langle N_{\mathrm{adj}}\rangle$ for $N=12$, showing that the collective-decay correlations persist over a broad cooperativity window.
  (c) Peak logarithmic negativity per site $\max_t E_{\mathcal N}/N$ versus inverse cooperativity for $N=6,8,10,12$; the dashed line marks $1/C=1$.
  (d)  $E_{\mathcal N}(t)$ for different common-mode dephasing values $\gamma_\phi$ (blue to yellow) for $N=10$, demonstrating collapse onto the lossless dynamics over $\gamma_\phi/\Gamma=0,\ldots,10^2$.
  Panels (a)--(c) use the adiabatically eliminated EAST-constrained spin model with unconstrained single-site loss, $\kappa=1$, and $\Delta=0$.
  Panel (d) uses the same EAST constraint with common-mode dephasing, $g=1$, $\kappa=40$, and $\Delta=0$.}
  \label{fig:endmatter_loss_dephasing}
\end{figure}

A possible experimental implementation can be obtained by extending the Rydberg antiblockade scheme proposed in Ref.~\cite{ValenciaTortoraEtAl2024} to a cavity-QED setting. While Ref.~\cite{ValenciaTortoraEtAl2024} considered directly driven two-level Rydberg atoms, here we employ a cavity-assisted Raman scheme involving three-level atoms collectively coupled to a single cavity mode. In the dispersive regime, adiabatic elimination of the intermediate level and accounting for Rydberg interaction-induced shifts yields an effective constrained spin-photon model of the form introduced in Eq.~(\ref{eq:H}).

We consider a collection of three-level atoms with ground state $\ket g$, intermediate state $\ket e$, and Rydberg state $\ket r$. The cavity mode couples the transition $\ket g\leftrightarrow\ket e$ with strength $g$, while a classical laser field drives the transition $\ket e\leftrightarrow\ket r$ with time- and position-dependent Rabi frequency $\Omega f_j(t)$. For large detuning between the cavity frequency and the transition frequency between the ground and intermediate state $\vert \omega_c - \omega_e \vert = |\Delta|\gg g,\Omega,\gamma_e$, the state $\ket e$ is only virtually populated and can therefore be adiabatically eliminated. The resulting dynamics in the reduced $\{\ket g,\ket r\}$ subspace are described by
\begin{align}
    H &= - \hbar \sum_j \delta^{\mathrm{eff}}_{j}(t) n_j + \hbar g_\mathrm{eff} \sum_j \left( f_j(t) a \sigma_j^{+} + f_j^*(t) a^\dagger \sigma_j^{-} \right) \nonumber \\
    &\qquad + \frac{1}{2} \sum_{j \neq l} V_{jl} n_j n_l \, .
\end{align}
where $\sigma_j^{-}$ accounts for the transition from the Rydberg state to the ground state, $n_j$ gives the occupation of the Rydberg state, and $\delta^{\mathrm{eff}}_{j}(t) = \delta + g^2 / \Delta a^\dagger a - \Omega^2 \vert f_j(t) \vert^2 / \Delta$ is a time- and position-dependent effective detuning due to AC Stark shifts coming from the cavity field as well as the classical drive and $\delta$ is an additional two-photon detuning. The off-resonant two-photon Raman process generates an effective atom-cavity coupling strength $g_\mathrm{eff} = g \Omega / \Delta$. The last term describes van der Waals interactions between Rydberg excitations, where $V_{jl} = C_6 / \vert r_j - r_l \vert^6$ for atoms located at positions $r_j$.

The kinetic constraints are implemented spectroscopically through interaction-induced Rydberg antiblockade shifts following Ref.~\cite{ValenciaTortoraEtAl2024}; see also Supplemental Material for details. The cavity-assisted Raman transition becomes resonant only when the two-photon laser detuning compensates the interaction shift generated by neighboring Rydberg excitations. Constraint-dependent resonances can therefore be engineered through the atomic geometry and the choice of laser detunings.

The AND and OR constraints are realized using atoms arranged with uniform spacing $d$, such that neighboring Rydberg excitations produce an interaction shift $V = C_6/d^6$. Driving all atoms uniformly and choosing $\delta = - 2V$ realizes the AND constraint, since the Raman transition becomes resonant only when both neighboring atoms are in the Rydberg state. Likewise, setting $\delta = -V$ realizes the OR constraint, for which resonance occurs whenever exactly one neighboring atom is excited.

To implement the EAST constraint, the atoms are instead arranged with staggered spacings such that the distance between odd-even pairs is $d_1$, while the distance between even-odd pairs is $d_2$. The corresponding interaction shifts are therefore given by $V_1 = C_6/d_1^6$ and $V_2 = C_6/d_2^6$, respectively. Two classical Raman drives with equal Rabi frequency $\Omega$ but distinct detunings are then applied. One drive is detuned by the nearest-neighbor interaction shift $V_{j-1,j}$ and becomes resonant only when the left neighbor of atom $j$ is in the Rydberg state. The second drive is detuned by $V_1 + V_2$ and compensates the additional interaction shift that arises when both neighboring atoms are excited, thereby maintaining resonance in this configuration as well. In the strong-interaction limit $V_1, V_2, \vert V_1 - V_2 \vert \gg g_\mathrm{eff}, \kappa$ all remaining off-resonant processes are suppressed, and the dynamics reduce to the kinetically constrained Hamiltonian
\begin{align}
    H &= - \hbar \sum_j \left( \delta + \frac{g^2}{\Delta} a^\dagger a - \frac{2 \Omega ^2}{\Delta} \right) n_j  \nonumber \\
    & \quad + g_\mathrm{eff} \sum_{j \geq 2} n_{j-1} ( a \sigma_j^{+} + a^\dagger \sigma_j^{-}) + \sum_{j} \sum_{l \geq j+2} V_{jl} n_j n_l,
\end{align}
where the last term accounting for residual interactions beyond nearest neighbors.

A key imperfection in this scheme is residual single-site decay. Virtual population of $\ket e$ induces unwanted decay from $\ket r$ to $\ket g$, described by local jump operators $L_j^{\mathrm{loss}}=\sqrt{\gamma_{\mathrm{eff}}}\,\sigma_j^-$ with $\gamma_{\mathrm{eff}}=\gamma_e(\Omega/\Delta)^2$. This competes with the collective decay rate $\Gamma=4g_{\mathrm{eff}}^2/\kappa=4g^2\Omega^2/(\Delta^2\kappa)$, giving
\begin{align}
    \frac{\gamma_{\mathrm{eff}}}{\Gamma}
    =
    \frac{\gamma_e \kappa}{4g^2}
    =
    \frac{1}{4C},
\end{align}
where $C=g^2/(\gamma_e\kappa)$ is the single-atom cooperativity. Thus single-site loss is controlled by the inverse cooperativity. Recent cavity-Rydberg experiments report $\gamma_e=2\pi\times 6\,\mathrm{MHz}$, $\kappa=2\pi\times 0.85\,\mathrm{MHz}$, and $g=2\pi\times 2.26\,\mathrm{MHz}$, corresponding to $C\simeq 1$~\cite{SantisRydbergCavity2026}.

For finite single-site loss, the unentangled empty state is the unique stationary state. Nevertheless, substantial transient entanglement survives. Figure~\ref{fig:endmatter_loss_dephasing}(a) shows the half-chain logarithmic negativity $E_{\mathcal N}(t)$ for $N=12$ and several cooperativities. At experimentally accessible $C\sim 1$, the peak reaches $62\%$ of its lossless value, showing that entanglement forms on the superradiant timescale before single-site loss dominates. Figure~\ref{fig:endmatter_loss_dephasing}(c) shows $\max_t E_{\mathcal N}/N$ versus inverse cooperativity for several system sizes. Although the negativity decreases with $1/C$, the curves remain close across the accessible sizes, indicating approximately extensive transient entanglement even near $C\sim 1$.

Finite loss also impacts the use of the adjacent excitation count $\langle N_{\mathrm{adj}}\rangle$ as an entanglement witness. Without loss, $\langle N_{\mathrm{adj}}\rangle$ develops a plateau associated with the dark manifold. With loss, the plateau eventually decays; see Fig.~\ref{fig:endmatter_loss_dephasing}(b). At large cooperativity, however, a long-lived shoulder remains, showing that the system stays near the dark manifold before loss takes over. Thus $\langle N_{\mathrm{adj}}\rangle$ is no longer an exact witness at finite loss, but its transient enhancement still signals dark-state correlations.

A second imperfection is common-mode dephasing, for example from laser phase noise. We model it by adding $\gamma_\phi \mathcal{D}[S^z]$ to the master equation. Figure~\ref{fig:endmatter_loss_dephasing}(d) shows that the logarithmic-negativity dynamics are essentially unchanged for $\gamma_\phi/\Gamma=0,\ldots,10^2$. The reason is simple: the dark manifold is an eigenspace of $S^z$. Common-mode dephasing therefore does not drive transitions out of this manifold and does not directly degrade the stationary-state entanglement. In the Supplemental Material~\cite{SI}, we discuss further imperfections, including next-nearest-neighbor Rydberg interactions and single-site dephasing, and show that the main results remain robust.

\end{document}


\title{Supplemental Material: Extensive mixed-state entanglement in kinetically constrained superradiance}

\date{\today}
\maketitle

\tableofcontents
\section{Adiabatic elimination of the cavity}

We adopt the conventions of the main-text model section. Define
\begin{equation}
F \equiv \sum_j P_j \sigma_j^-,
\end{equation}
where the constraint operators \(P_j\) commute with all \(\sigma_\ell^z\) (equivalently all \(n_\ell\)). The Hamiltonian is
\begin{align}
H_0 &= \omega_c a^\dagger a + \omega_s S^z, \qquad
S^z = \frac12 \sum_j \sigma_j^z, \\
H_{\mathrm{int}} &= g(a^\dagger F + a F^\dagger),
\end{align}
and cavity loss is described by \(\kappa\,\mathcal D[a](\rho)\), with
\begin{equation}
\mathcal D[L](\rho)=L\rho L^\dagger-\tfrac12\{L^\dagger L,\rho\}.
\end{equation}

Moving to the interaction picture with respect to \(H_0\), one has \(a(t)=e^{-i\omega_c t}a\). Since the \(P_j\) commute with \(S^z\), while \([S^z,\sigma_j^-]=-\sigma_j^-\), it follows that \([H_0,F]=-\omega_s F\), hence \(F(t)=e^{-i\omega_s t}F\). Therefore
\begin{equation}
V_I(t)=g\left(a^\dagger F\,e^{i\Delta t}+aF^\dagger e^{-i\Delta t}\right),
\qquad
\Delta=\omega_c-\omega_s.
\end{equation}

In the bad-cavity regime the cavity remains close to vacuum, so we use the Born approximation
\begin{equation}
\rho(t)\approx \rho_s(t)\otimes |0\rangle\langle 0|.
\end{equation}
To second order in \(g\), the standard Born--Markov expansion gives
\begin{equation}
\dot\rho_s(t)
=
-\int_0^\infty d\tau\;
\mathrm{Tr}_{\mathrm{cav}}
\Big(
[V_I(\tau),[V_I(0),\rho_s(t)\otimes |0\rangle\langle 0|]]
\Big)
+\mathcal O(g^3).
\end{equation}
Since the cavity is in vacuum, the only nonvanishing correlator is
\begin{equation}
\langle a(\tau)a^\dagger(0)\rangle = e^{-(\kappa/2+i\Delta)\tau},
\qquad \tau\ge 0.
\end{equation}
All other two-time correlators vanish. Evaluating the kernel then yields
\begin{align}
\dot\rho_s
&=
\alpha\big(F\rho_s F^\dagger - F^\dagger F\,\rho_s\big)
+\alpha^*\big(F\rho_s F^\dagger - \rho_s F^\dagger F\big)
+\mathcal O(g^3),
\end{align}
with
\begin{equation}
\alpha
=
g^2\int_0^\infty d\tau\, e^{-(\kappa/2+i\Delta)\tau}
=
\frac{g^2}{\kappa/2+i\Delta}.
\end{equation}
Writing
\begin{equation}
\Gamma = 2\,\mathrm{Re}\,\alpha
= \frac{g^2\kappa}{\Delta^2+(\kappa/2)^2},
\qquad
\chi = \mathrm{Im}\,\alpha
= -\frac{g^2\Delta}{\Delta^2+(\kappa/2)^2},
\end{equation}
the effective spin master equation becomes
\begin{equation}
\dot\rho_s
=
-i[\chi F^\dagger F,\rho_s]
+
\Gamma\,\mathcal D[F](\rho_s)
+
\mathcal O(g^3).
\end{equation}

Thus the cavity elimination has the same form for any constrained jump operator
\(
F=\sum_j P_j \sigma_j^-
\)
whose constraint operators are diagonal in the \(z\) basis: the only effect is that the collective jump operator \(F\) is dressed by the chosen local constraints.

The elimination is controlled by the smallness of the cavity-mediated scale compared to the detuning and linewidth, i.e.
\begin{equation}
\Delta^2+\kappa^2 \gg g^2 N,
\end{equation}
so that cavity excitations remain virtual and higher-order corrections are suppressed.

\section{Equations of motion}

\label{sec:ode_pq_derivation}
We consider the spin-only description obtained after eliminating the cavity mode in the bad-cavity limit. The reduced master equation is
\begin{equation}
\dot\rho = -i\chi\,[F^\dagger F,\rho]
+\Gamma\Big(F\rho F^\dagger-\tfrac12\{F^\dagger F,\rho\}\Big),
\label{eq:ME}
\end{equation}
with the constrained collective lowering operator
\begin{equation}
F=\sum_j P_j\,\sigma_j^-.
\label{eq:F_general}
\end{equation}
Here $P_j$ denotes an arbitrary local density constraint acting on the neighborhood of site $j$. The only property we require is that it is diagonal in the $\sigma^z$ basis, i.e.
\begin{equation}
[P_i,\sigma_j^z]=0
\qquad\text{for all }i,j,
\label{eq:P_diag}
\end{equation}
or equivalently that they commute with all local densities $n_j=(1+\sigma_j^z)/2$. Throughout this note we set the detuning to zero, $\chi=0$, so that the dynamics is purely dissipative. We assume periodic boundary conditions and, when needed later, translation-invariant initial conditions.

In the Heisenberg picture, the adjoint Lindblad generator reads
\begin{equation}
\frac{dO}{dt}=\Gamma\,\mathcal{D}^\dagger_F(O),
\qquad
\mathcal{D}^\dagger_F(O)=F^\dagger O F-\tfrac12\{F^\dagger F,O\}.
\label{eq:adjoint}
\end{equation}
A useful identity is
\begin{equation}
\mathcal{D}^\dagger_F(O)
=\tfrac12\Big(F^\dagger[O,F]+[F^\dagger,O]F\Big),
\label{eq:adjoint_id}
\end{equation}
which follows by expanding the commutators.

The key collective observable is
\begin{equation}
S^z=\frac12\sum_j \sigma_j^z.
\end{equation}
The total emission rate (superradiant intensity) is proportional to $\langle F^\dagger F\rangle$. Since each constraint $P_j$ commutes with $S^z$, while
\begin{equation}
[S^z,\sigma_j^-]=-\sigma_j^-,
\end{equation}
we obtain
\begin{align}
[S^z,F]
&=\sum_j P_j\,[S^z,\sigma_j^-] \nonumber\\
&=-\sum_j P_j\,\sigma_j^- \nonumber\\
&=-F,
\label{eq:SzFcomm}
\end{align}
and similarly
\begin{equation}
[F^\dagger,S^z]=-F^\dagger.
\label{eq:FdagSzcomm}
\end{equation}
Inserting this into Eq.~\eqref{eq:adjoint_id} with $O=S^z$ gives the exact operator identity
\begin{align}
\mathcal{D}_F^\dagger(S^z)
&=\tfrac12\Big(F^\dagger[S^z,F]+[F^\dagger,S^z]F\Big)\nonumber\\
&=\tfrac12\Big(F^\dagger(-F)+(-F^\dagger)F\Big)\nonumber\\
&=-F^\dagger F.
\label{eq:DdaggerSz}
\end{align}
Therefore,
\begin{equation}
\boxed{\frac{d}{dt}\langle S^z\rangle=-\Gamma\,\langle F^\dagger F\rangle.}
\label{eq:Sz_exact}
\end{equation}

Equation \eqref{eq:Sz_exact} is exact and holds for any constrained jump operator of the form \eqref{eq:F_general}, provided the local constraints are diagonal in the $\sigma^z$ basis. Thus, independently of the detailed choice of kinetic constraint, the population dynamics is controlled by the expectation value $\langle F^\dagger F\rangle$. The remaining task in deriving closed equations of motion is therefore to express or approximate this quantity in terms of a reduced set of observables appropriate to the chosen constraint.

\subsection{Nearest-neighbor AND constraint as a concrete scaling example}
\label{app:and_scaling_example}

We first give an explicit calculation for one local rule, which will also serve as a useful example of the general bounds below.  Take the nearest-neighbor AND constraint
\begin{equation}
F_{\rm AND}
=
\sum_{j=1}^N n_{j-1}\sigma_j^- n_{j+1},
\qquad
n_j=\frac{1+\sigma_j^z}{2},
\label{eq:F_AND_def}
\end{equation}
with periodic boundary conditions.  Starting from
\begin{equation}
\ket{\Uparrow}=\ket{\uparrow\uparrow\cdots\uparrow},
\qquad
\ket{\psi_k}=F_{\rm AND}^k\ket{\Uparrow},
\end{equation}
the layer-$k$ intensity is
\begin{equation}
\langle F_{\rm AND}^\dagger F_{\rm AND}\rangle_k
=
\frac{\bra{\psi_k}F_{\rm AND}^\dagger F_{\rm AND}\ket{\psi_k}}
{\braket{\psi_k|\psi_k}}
=
\frac{\|\psi_{k+1}\|^2}{\|\psi_k\|^2}.
\label{eq:FAND_normratio}
\end{equation}
After a site has flipped down, neither of its nearest neighbors can flip later.  Thus the reachable configurations after $k$ jumps are precisely the sets of $k$ down spins with no adjacent down spins.  Every such set is reachable, and all $k!$ flip orders contribute with the same amplitude.  The number of such configurations on a ring is
\begin{equation}
C_{N,k}^{\rm AND}
=
\frac{N}{N-k}\binom{N-k}{k},
\qquad
0\le k\le \left\lfloor\frac{N}{2}\right\rfloor .
\label{eq:C_Nk_AND}
\end{equation}
Therefore
\begin{equation}
\|\psi_k\|^2=(k!)^2 C_{N,k}^{\rm AND},
\qquad
\langle F_{\rm AND}^\dagger F_{\rm AND}\rangle_k
=(k+1)^2\frac{C_{N,k+1}^{\rm AND}}{C_{N,k}^{\rm AND}},
\label{eq:FAND_count_ratio}
\end{equation}
or explicitly
\begin{equation}
\boxed{
\langle F_{\rm AND}^\dagger F_{\rm AND}\rangle_k
=
(k+1)\frac{(N-2k)(N-2k-1)}{N-k-1}.
}
\label{eq:FAND_exact}
\end{equation}

\begin{figure}[t]
  \centering
  \includegraphics[width=0.92\linewidth]{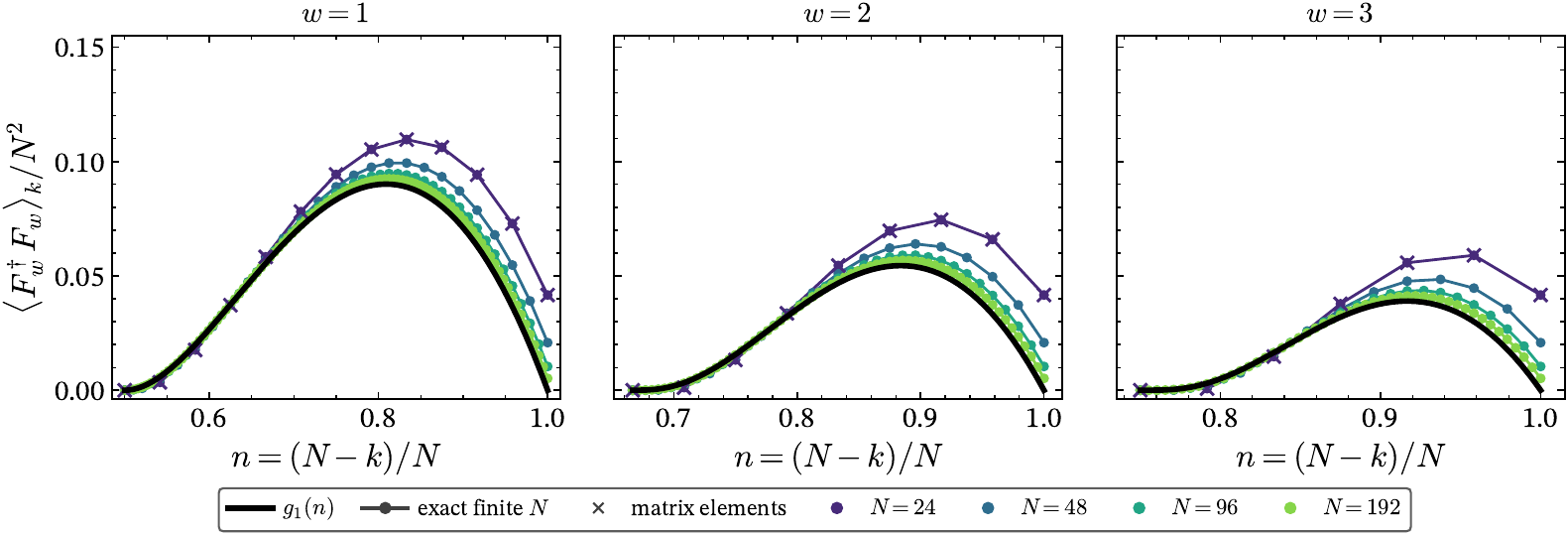}
  \caption{\textbf{Finite-size checks and thermodynamic scaling of finite-range AND constraints.}
  Dots show $\langle F_w^\dagger F_w\rangle_k/N^2$ obtained by explicit sparse application of the range-$w$ AND operator $F_w=\sum_j(\prod_{d=1}^w n_{j-d})\sigma_j^-(\prod_{d=1}^w n_{j+d})$ to $\ket{\Uparrow}$, while crosses show the corresponding exact finite-$N$ counting expression.  Colors distinguish system sizes and panels show representative constraint ranges $w=1,2,3$.  The $w=1$ panel is the nearest-neighbor AND model of Eq.~\eqref{eq:F_AND_def}, with thermodynamic scaling function Eq.~\eqref{eq:gAND_n}.}
  \label{fig:range_w_fdagf_scaling}
\end{figure}

Set $\nu=k/N$ and $n=1-\nu$.  For fixed $\nu\in[0,1/2]$,
\begin{equation}
\frac{1}{N^2}\langle F_{\rm AND}^\dagger F_{\rm AND}\rangle_k
\xrightarrow[N\to\infty]{}
g_{\rm AND}(\nu)
=
\nu\,\frac{(1-2\nu)^2}{1-\nu}.
\label{eq:gAND_nu}
\end{equation}
Equivalently,
\begin{equation}
\boxed{
g_{\rm AND}(n)
=
(1-n)\frac{(2n-1)^2}{n},
\qquad
n\in\left[\frac12,1\right].
}
\label{eq:gAND_n}
\end{equation}
Outside this interval the scaling function vanishes because such densities are inaccessible along this constrained trajectory. In Fig. \ref{fig:range_w_fdagf_scaling} we compare the asymptotic scaling function \eqref{eq:gAND_n} to the finite system size exact analytical form \eqref{eq:FAND_exact} and numerical evaluation of \eqref{eq:FAND_normratio} confirming exact matching. 

This concrete model already shows the essential scaling.  The upper bound is immediate from $\|F_{\rm AND}\|\le N$:
\begin{equation}
\langle F_{\rm AND}^\dagger F_{\rm AND}\rangle_k\le \|F_{\rm AND}\|^2\le N^2.
\end{equation}
For a lower bound, choose $\nu_0=1/4$.  Then
\begin{equation}
g_{\rm AND}(\nu_0)=\frac{1}{12}>0,
\label{eq:gAND_positive_bound}
\end{equation}
and thus for any sequence $k_N/N\to\nu_0$,
\begin{equation}
\langle F_{\rm AND}^\dagger F_{\rm AND}\rangle_{k_N}
=\frac{N^2}{12}+o(N^2).
\end{equation}
Combining the upper and lower bounds gives
\begin{equation}
\boxed{
\max_k\langle F_{\rm AND}^\dagger F_{\rm AND}\rangle_k=\Theta(N^2).
}
\label{eq:FAND_quadratic_scaling}
\end{equation}

The corresponding thermodynamic rate equation for the excitation density is
\begin{equation}
\frac{dn}{dt}=-\frac{\Gamma}{N}\langle F_{\rm AND}^\dagger F_{\rm AND}\rangle
\simeq -\Gamma N\,g_{\rm AND}(n),
\qquad
\tau:=\Gamma Nt,
\qquad
\frac{dn}{d\tau}=-g_{\rm AND}(n).
\label{eq:FAND_ode_rescaled}
\end{equation}
For an initial condition $n(0)=n_0$,
\begin{equation}
\tau(n)=\int_n^{n_0}\frac{dn'}{g_{\rm AND}(n')}.
\label{eq:FAND_solution_quadrature}
\end{equation}
Starting from $\ket{\Uparrow}$, the continuum integral is cut off after the first jump at $n_0=1-1/N$. Since $g_{\rm AND}(n)\sim 1-n$ as $n\to1$, the time to reach the maximizer $n_\star$ of $g_{\rm AND}(n)$ has the universal leading form
\begin{equation}
t_\star(N)=\frac{\ln N+C_\star+o(1)}{\Gamma N},
\label{eq:t_star_log_over_N_AND}
\end{equation}
with a nonuniversal constant $C_\star$ set by the regular part of the integral in Eq.~\eqref{eq:FAND_solution_quadrature}.

\subsection{Lower bound for all local Boolean constraints}
\label{app:boolean_no_bottleneck_bound}

While the AND model has both the superradiant peak intensity and time scaling, this actually holds for all local Boolean constraints. We will now derive a lower bound valid for all constraints with width $w$. Consider a ring of $N$ spins and the constrained jump operator
\begin{equation}
F_f=\sum_{j=1}^N P_j^{(f)}\sigma_j^- ,
\label{eq:Ff_boolean}
\end{equation}
where the superscript \(f\) labels the Boolean rule. Each \(P_j^{(f)}\) is diagonal in the computational basis, takes values in $\{0,1\}$, and depends only on the occupations of the $2w$ neighboring sites
\begin{equation}
j-w,\dots,j-1,j+1,\dots,j+w .
\end{equation}
The range $w$ is fixed as $N\to\infty$.  We assume only  \begin{equation}
P_j^{(f)}(1,\dots,1)=1,
\label{eq:boolean_normalization}
\end{equation} otherwise the fully inverted state cannot decay. Starting from $\ket{\Uparrow}$, define
\begin{equation}
\ket{\psi_k}=F_f^k\ket{\Uparrow},
\qquad
B_k=\|\psi_k\|^2,
\qquad
\left\langle F_f^\dagger F_f\right\rangle_k
=\frac{B_{k+1}}{B_k}
\label{eq:layer_rate_boolean}
\end{equation}
We can express the state $\ket{\psi_k}$ in the computational basis. We denote $\ket{S}$ a computational basis state defined by which spins have flipped from the fully polarized state. This means $S =\{j_1, \dots, j_k\}$ is a set of decayed sites with size $|S|=k$. We can now expand the states $\ket{\psi_k}$ in the computational basis defined by $S$ with the basis coefficients $a_f(S)$
\begin{equation}
\ket{\psi_k}
=
\sum_{|S|=k} a_f(S)\ket{S},
\qquad
B_k
=
\sum_{|S|=k} a_f(S)^2 .
\label{eq:Bk_boolean_history}
\end{equation}

Note, as in the last section, $a_f(S)$ is given by the number of decay histories leading to the basis state $S$. A decay history is a tuple indicating in which order the spins decayed. For example, a state $S=\{j_1, j_2, \dots, j_k\}$ might be reached via the decay histories $(j_1, j_2, \dots, j_k)$ and $(j_2, j_1, \dots, j_k)$.  In the unconstrained Dicke model arbitrary permutations of the set $S$ are admissible decay histories. This is because the model is permutationally symmetric, so the decay order does not matter. Constraints generally disallow some decay histories. 

Our goal now is to find a lower bound for the emission rate 
\begin{align}
    \braket{F_f^\dagger F_f}_k = \frac{B_{k+1}}{B_k} = \frac{\sum_{|S'|=k+1} a_f^2(S')}{\sum_{|S|=k}a_f^2(S)}\ ,
\end{align}
independent of constraint. The core idea is that all decay histories after $k+1$ emissions can be constructed from the decay histories after $k$ emissions via insertion of one additional decay site. However, here we face the issue that some insertions might be disallowed by the constraints. The emission order generally does not commute for this problem. 
  
Nevertheless, as the constraint is finite range, there is a set of sites that is sufficiently far away from any already decayed spin in $S$ so it is  unconstrained.   Given a range $w$ constraint, we can define this set of unconstrained spins as 
\begin{equation}
G_w(S)
=
\{j\notin S:\dist(j,S)>w\}.
\label{eq:G_boolean_def}
\end{equation}
For $S=\emptyset$, we use the convention $G_w(\emptyset)=\{1,\dots,N\}$. Let us formalize the fact that we obtain an admissible decay history given a decay history ending at $S$ via insertion of a spin in $G_w(S)$. 

\begin{lemma}[Insertion of an isolated decay]
\label{lem:boolean_insertion}
Let $h$ be a valid ordered decay history ending at a set $S$ with $|S|=k$, and let $j\in G_w(S)$.  Then inserting the decay of $j$ at any of the $k+1$ positions in $h$ gives a valid ordered decay history ending at $S\cup\{j\}$.
\end{lemma}

\begin{proof}
At the instant when the inserted decay of $j$ occurs, the already decayed spins form a subset of $S$.  Since $j\in G_w(S)$, none of these spins lies in the constraint neighborhood of $j$.  Therefore all $2w$ neighbors relevant to \(P_j^{(f)}\) are still up, and Eq.~\eqref{eq:boolean_normalization} allows the decay of $j$.  The inserted spin is also outside the constraint neighborhood of every original decay site in $S$, so the inserted step does not spoil any step of the original history.
\end{proof}
Using the ability to insert decay sites we can now derive the following lower bound. 

\begin{theorem}
\label{thm:boolean_no_bottleneck}
For every $k\ge0$ satisfying
\begin{equation}
N-(2w+1)k>0,
\label{eq:boolean_positive_window}
\end{equation}
one has $B_k>0$ and
\begin{equation}
\left\langle F_f^\dagger F_f\right\rangle_k
\ge
(k+1)\bigl[N-(2w+1)k\bigr].
\label{eq:boolean_no_bottleneck_intensity}
\end{equation}

\end{theorem}
\begin{proof}
We bound the layer intensity directly from
\begin{equation}
\left\langle F_f^\dagger F_f\right\rangle_k
=
\frac{B_{k+1}}{B_k}
=
\frac{\sum_{|T|=k+1} a_f(T)^2}
{\sum_{|S|=k} a_f(S)^2}.
\label{eq:boolean_ratio_start}
\end{equation}

Fix a reachable set $S$ with $|S|=k$ and $a_f(S)>0$, and choose an
isolated site $j\in G_w(S)$.  By Lemma~\ref{lem:boolean_insertion},
the decay of $j$ may be inserted into any valid history ending at $S$
at any of the $k+1$ possible positions.  Therefore the extended set
\begin{equation}
T=S\cup\{j\}
\end{equation}
has at least $(k+1)$ times as many valid histories:
\begin{equation}
a_f(S\cup\{j\})
\ge
(k+1)a_f(S).
\label{eq:boolean_coeff_extension}
\end{equation}
Squaring gives
\begin{equation}
a_f(S\cup\{j\})^2
\ge
(k+1)^2 a_f(S)^2.
\label{eq:boolean_coeff_extension_squared}
\end{equation}

We now use these isolated extensions to lower bound the numerator
$B_{k+1}$.  Summing over all pairs $(S,j)$ with $|S|=k$ and
$j\in G_w(S)$ gives
\begin{equation}
\sum_{|S|=k}\sum_{j\in G_w(S)}
a_f(S\cup\{j\})^2 .
\end{equation}
This sum may count a given $(k+1)$-element set $T$ more than once,
because $T$ can arise from deleting different sites $j\in T$.
However, there are at most $k+1$ such choices.  Hence
\begin{equation}
B_{k+1}
=
\sum_{|T|=k+1}a_f(T)^2
\ge
\frac{1}{k+1}
\sum_{|S|=k}\sum_{j\in G_w(S)}
a_f(S\cup\{j\})^2 .
\label{eq:boolean_Bkp1_direct_lower}
\end{equation}
Using Eq.~\eqref{eq:boolean_coeff_extension_squared},
\begin{align}
B_{k+1}
&\ge
\frac{1}{k+1}
\sum_{|S|=k}\sum_{j\in G_w(S)}
(k+1)^2 a_f(S)^2
\nonumber\\
&=
(k+1)
\sum_{|S|=k}
|G_w(S)|\,a_f(S)^2 .
\label{eq:boolean_Bkp1_lower_final}
\end{align}
Dividing by $B_k$
yields
\begin{equation}
\left\langle F_f^\dagger F_f\right\rangle_k
\ge
(k+1)
\frac{
\sum_{|S|=k}|G_w(S)|\,a_f(S)^2
}{
\sum_{|S|=k}a_f(S)^2
}.
\label{eq:boolean_weighted_G_bound}
\end{equation}

Finally, each decayed site excludes at most itself and its $2w$
neighboring sites from $G_w(S)$, so
\begin{equation}
|G_w(S)|
\ge
N-(2w+1)k .
\label{eq:boolean_G_lower_bound}
\end{equation}
Substituting this into Eq.~\eqref{eq:boolean_weighted_G_bound} gives
\begin{equation}
\left\langle F_f^\dagger F_f\right\rangle_k
\ge
(k+1)\bigl[N-(2w+1)k\bigr],
\end{equation}
which is Eq.~\eqref{eq:boolean_no_bottleneck_intensity}.

Since $B_0=1$, and the right-hand side is strictly positive whenever
$N-(2w+1)k>0$, the same argument also implies inductively that
$B_k>0$ throughout this window.
\end{proof}

\paragraph*{Quadratic intensity and waiting time.}
Theorem~\ref{thm:boolean_no_bottleneck} implies that the constrained collective decay has a genuinely quadratic intensity at an extensive number of emitted photons.  Let $c=2w+1$.  For any fixed $\nu = k/N$ with $0<\nu<1/c$ and $k=\lfloor \nu N\rfloor$,
\begin{equation}
\left\langle F_f^\dagger F_f\right\rangle_k
\ge
(k+1)(N-ck)
=
\nu(1-c\nu)N^2+o(N^2).
\label{eq:boolean_quadratic_lower}
\end{equation}
Conversely, the trivial operator-norm bound
\begin{equation}
\|F_f\|
\le
\sum_{j=1}^N \|P_j^{(f)}\sigma_j^-\|
\le
N
\end{equation}
implies $\left\langle F_f^\dagger F_f\right\rangle_k\le N^2$ whenever $B_k>0$.  Hence
\begin{equation}
\max_k \left\langle F_f^\dagger F_f\right\rangle_k=\Theta(N^2)
\label{eq:boolean_peak_theta}
\end{equation}
for every fixed-range Boolean constraint satisfying Eq.~\eqref{eq:boolean_normalization}.  The constraint can change the prefactor and the late-time decay structure, but it cannot remove the early collective $N^2$ burst.

The same bound controls the traversal time through the early jump layers.  In the click-limit description, the normalized state in the $k$th layer has instantaneous jump intensity
\begin{equation}
\left\langle F_f^\dagger F_f\right\rangle_k
=
\frac{\|F_f\psi_k\|^2}{\|\psi_k\|^2}
=
\frac{B_{k+1}}{B_k}.
\end{equation}
Thus the typical waiting time for the transition from layer $k$ to layer $k+1$ is
\begin{equation}
\tau_k\sim \frac{1}{\Gamma\left\langle F_f^\dagger F_f\right\rangle_k} .
\end{equation}
For $K=\lfloor \nu N\rfloor$ with $\nu<1/c$, Theorem~\ref{thm:boolean_no_bottleneck} gives
\begin{equation}
\sum_{k=0}^{K-1}\tau_k
\lesssim
\frac{1}{\Gamma(1-c\nu)N}
\sum_{k=0}^{K-1}\frac{1}{k+1}
=
O\!\left(\frac{\log N}{\Gamma N}\right).
\label{eq:boolean_waiting_time_bound}
\end{equation}
The logarithm comes from summing the earliest waiting times,
\begin{equation}
\frac{1}{\Gamma N},\quad
\frac{1}{2\Gamma N},\quad
\frac{1}{3\Gamma N},\quad \dots,
\end{equation}
rather than from a bottleneck at finite density.  At any fixed density below $1/(2w+1)$, the intensity itself is already of order $N^2$.

\section{Construction of dark states for the open EAST loss operator}
\label{app:east_dark_states}

In this Appendix we construct the dark states of the open EAST loss operator
\begin{equation}
F_{\rm open}=\sum_{j=2}^{N} n_{j-1}\sigma_j^-,
\qquad
n_j=\frac{1+\sigma_j^z}{2},
\label{eq:EastL}
\end{equation}
where the first spin is unconstrained. Dark states are the zero modes of
$F_{\rm open}^\dagger F_{\rm open}$. Since $F_{\rm open}^\dagger F_{\rm open}$ is positive semidefinite, they are
equivalently the states satisfying
\begin{equation}
F_{\rm open}|\psi\rangle=0.
\label{eq:dark_condition}
\end{equation}
The main-text numerics use periodic boundaries; we use the open chain here to
make the recursive cancellation packets local and notationally simple. The same
bulk packets embed in the periodic problem whenever their support does not wrap
around the boundary, while periodicity only adds the corresponding boundary
cancellation conditions.

Our numerical data separate the dark states into three coarse groups using
\begin{equation}
N_{\rm adj}=\sum_{j=2}^{N} n_{j-1}n_j,
\qquad
N_{\rm tri}=\sum_{j=2}^{N-1} n_{j-1}n_j n_{j+1}.
\label{eq:NadjNtri}
\end{equation}
The first group has $N_{\rm adj}=N_{\rm tri}=0$, the second has
$N_{\rm adj}\neq 0$ but $N_{\rm tri}=0$, and the third has
$N_{\rm tri}\neq 0$. We show below that these are the first three layers of a
natural recursive hierarchy of dark states.

\subsection{Kernel condition in the computational basis}

Let
\begin{equation}
|\mathbf n\rangle = |n_1n_2\cdots n_N\rangle,
\qquad n_j\in\{0,1\},
\end{equation}
denote a computational-basis state, and let $\mathbf e_j$ be the bitstring with
a single $1$ at site $j$. Acting with $F_{\rm open}$ on a basis state gives
\begin{equation}
F_{\rm open}|\mathbf n\rangle
=
\sum_{j=2}^{N} n_{j-1}n_j\, |\mathbf n-\mathbf e_j\rangle.
\label{eq:L_on_basis}
\end{equation}
Thus every adjacent motif $11$ contributes one decay process by flipping the
\emph{right} excitation to zero.

For a general state
\begin{equation}
|\psi\rangle=\sum_{\mathbf n} c_{\mathbf n}|\mathbf n\rangle,
\end{equation}
the dark-state condition \eqref{eq:dark_condition} becomes
\begin{equation}
\forall\,\mathbf m:\qquad
\sum_{j:\,m_{j-1}=1,\;m_j=0} c_{\mathbf m+\mathbf e_j}=0.
\label{eq:kernel_condition}
\end{equation}
Equation~\eqref{eq:kernel_condition} is the fundamental cancellation rule: for
every child configuration $|\mathbf m\rangle$, the amplitudes of all its
``parents'' must sum to zero.

It is convenient to introduce the higher block correlators
\begin{equation}
N_\ell=\sum_{j=2}^{N-\ell+2}\prod_{a=0}^{\ell-1} n_{j+a},
\label{eq:Nell}
\end{equation}
so that $N_2=N_{\rm adj}$ and $N_3=N_{\rm tri}$. These correlators detect
contiguous blocks of $\ell$ excitations and organize the dark states by the
largest run of $1$'s appearing in their support.

\subsection{First family: independent-set dark states}

The first family consists of basis states with no adjacent excitations. If
$n_{j-1}n_j=0$ for all $j$, then every term in $F_{\rm open}$ annihilates the state, so it
is dark. Hence
\begin{equation}
\mathcal K^{(1)}
=
\mathrm{span}\Big\{
|\mathbf n\rangle:\;
n_{j-1}n_j=0\ \forall j
\Big\}.
\label{eq:K1}
\end{equation}
These are precisely the product states with
\begin{equation}
N_{\rm adj}=N_{\rm tri}=0.
\end{equation}
Equivalently, $\mathcal K^{(1)}$ is the span of all independent-set bitstrings.

\subsection{Second family: one-dimer cancellation packets}

Let $|r\rangle\in\mathcal K^{(1)}$ be an independent-set root configuration.
Define the set of \emph{facilitable zeros}
\begin{equation}
S_1(r)=
\Big\{
i\in\{2,\dots,N\}:\;
r_{i-1}=1,\;
r_i=0,\;
(i=N\ \text{or}\ r_{i+1}=0)
\Big\}.
\label{eq:S1}
\end{equation}
For $i\in S_1(r)$, the state $\sigma_i^+|r\rangle$ contains a single adjacent
pair but no triple, and
\begin{equation}
F_{\rm open}\,\sigma_i^+|r\rangle=|r\rangle.
\label{eq:L_on_one_dimer}
\end{equation}
Therefore any zero-sum superposition over such parents is dark:
\begin{equation}
|D^{(2)}[r;\{\alpha_i\}]\rangle
=
\sum_{i\in S_1(r)} \alpha_i\,\sigma_i^+|r\rangle,
\qquad
\sum_{i\in S_1(r)}\alpha_i=0.
\label{eq:family2_general}
\end{equation}
A convenient basis is given by pairwise differences,
\begin{equation}
|D^{(2)}_{ij}[r]\rangle
=
(\sigma_i^+-\sigma_j^+)|r\rangle,
\qquad
i,j\in S_1(r).
\label{eq:family2_pair}
\end{equation}
Every state in this family satisfies
\begin{equation}
N_{\rm adj}\neq 0,
\qquad
N_{\rm tri}=0.
\end{equation}

As a simple example, for
\begin{equation}
|r\rangle=|10010\rangle,
\end{equation}
the facilitable zeros are $i=2$ and $j=5$, and
\begin{equation}
|D^{(2)}_{25}[r]\rangle
=
|11010\rangle-|10011\rangle.
\end{equation}
Indeed,
\begin{equation}
F_{\rm open}|11010\rangle=|10010\rangle,
\qquad
F_{\rm open}|10011\rangle=|10010\rangle,
\end{equation}
so the two contributions cancel.

\subsection{Third family: triple-support cancellation packets}

The first genuinely new states with $N_{\rm tri}\neq 0$ arise from local
configurations containing a block $111$. Let $|r\rangle\in\mathcal K^{(1)}$ and
define the set of \emph{extendable facilitable zeros}
\begin{equation}
S_2(r)=
\Big\{
i\in\{2,\dots,N-2\}:\;
r_{i-1}r_i r_{i+1}r_{i+2}=1000
\Big\}.
\label{eq:S2}
\end{equation}
For $i\in S_2(r)$, the state $\sigma_i^+\sigma_{i+1}^+|r\rangle$ contains a
local block $111$ but no block $1111$.

Take two sites $i,j\in S_2(r)$ with disjoint local windows. Then the following
state is dark:
\begin{equation}
\begin{aligned}
|D^{(3)}_{ij}[r]\rangle
={}&
\Big(
\sigma_i^+\sigma_{i+1}^+
+\sigma_j^+\sigma_{j+1}^+
-\sigma_i^+\sigma_{j+1}^+ \\
&\hspace{1.3cm}
-\sigma_{i+1}^+\sigma_j^+
-\sigma_i^+\sigma_j^+
\Big)|r\rangle.
\end{aligned}
\label{eq:family3_packet}
\end{equation}
The proof is local. Using that $|r\rangle$ has no adjacent excitations one
finds
\begin{align}
F_{\rm open}\,\sigma_i^+\sigma_{i+1}^+|r\rangle
&=
\sigma_i^+|r\rangle+\sigma_{i+1}^+|r\rangle,
\\
F_{\rm open}\,\sigma_j^+\sigma_{j+1}^+|r\rangle
&=
\sigma_j^+|r\rangle+\sigma_{j+1}^+|r\rangle,
\\
F_{\rm open}\,\sigma_i^+\sigma_{j+1}^+|r\rangle
&=
\sigma_{j+1}^+|r\rangle,
\\
F_{\rm open}\,\sigma_{i+1}^+\sigma_j^+|r\rangle
&=
\sigma_{i+1}^+|r\rangle,
\\
F_{\rm open}\,\sigma_i^+\sigma_j^+|r\rangle
&=
\sigma_i^+|r\rangle+\sigma_j^+|r\rangle.
\end{align}
Summing these contributions with coefficients $(+,+,-,-,-)$ gives
\begin{equation}
F_{\rm open}|D^{(3)}_{ij}[r]\rangle=0.
\end{equation}
States of this type have
\begin{equation}
N_{\rm tri}\neq 0.
\end{equation}

An explicit example is obtained from
\begin{equation}
|r\rangle=|1000100\rangle,
\qquad
i=2,\quad j=6,
\end{equation}
for which
\begin{equation}
\begin{aligned}
|D^{(3)}_{26}[r]\rangle
={}&
|1110100\rangle
+|1000111\rangle
-|1100101\rangle \\
&-|1010110\rangle
-|1100110\rangle.
\end{aligned}
\label{eq:Omega3_example}
\end{equation}
Direct evaluation confirms that $F_{\rm open}|D^{(3)}_{26}[r]\rangle=0$.

\subsection{Recursive hierarchy and general construction}

The three families above are only the first three layers of a broader
hierarchy. The correct organizing principle is the maximal run length of
consecutive excitations appearing in the support of a dark state.

For $m\ge 2$, define the local \emph{seed}
\begin{equation}
t_m = 1^m\,0\,1\,0^{m-1}.
\label{eq:tm_seed}
\end{equation}
Starting from $|t_m\rangle$, one constructs a dark state recursively as follows.

\begin{enumerate}
\item Assign coefficient $+1$ to the seed $|t_m\rangle$.
\item Act with $F_{\rm open}$. Every child produced by \eqref{eq:L_on_basis} must be
canceled.
\item For each uncanceled child, add all other parents required by the kernel
condition \eqref{eq:kernel_condition}.
\item Repeat until all children cancel.
\end{enumerate}

Because each application of $F_{\rm open}$ lowers the excitation number by one, this
procedure is triangular and terminates after finitely many steps. It defines a
local cancellation packet $\Omega_m$ whose support contains configurations with
maximal run length $m$ but no longer block.

The first nontrivial packets are
\begin{align}
\Omega_2
&=
|11010\rangle-|10011\rangle,
\label{eq:Omega2}
\\[1mm]
\Omega_3
&=
|1110100\rangle+|1000111\rangle-|1100101\rangle
\nonumber\\
&\quad
-|1010110\rangle-|1100110\rangle.
\label{eq:Omega3}
\end{align}
The packet $\Omega_4$ is obtained from the seed $111101000$ by the same
recursive cancellation procedure. It is the first example beyond the
$(N_{\rm adj},N_{\rm tri})$ classification, showing that $N_{\rm tri}\neq 0$
does not by itself distinguish all higher dark-state sectors.

The general structure of the kernel is therefore:
\begin{equation}
\ker F_{\rm open}
=
\mathrm{span}
\Big\{
\text{independent-set backgrounds decorated by embedded packets }
\Omega_m
\Big\}.
\label{eq:kernel_packet_structure}
\end{equation}
Packets can be inserted into an independent-set root provided their local
supports do not interfere. Products of packets on disjoint windows remain dark,
since the cancellations enforced by \eqref{eq:kernel_condition} are local.

In this language, the three groups seen numerically are:
\begin{align}
\text{Group I:} \quad & N_{\rm adj}=N_{\rm tri}=0
\quad\Longleftrightarrow\quad \mathcal K^{(1)},
\\
\text{Group II:} \quad & N_{\rm adj}\neq 0,\; N_{\rm tri}=0
\quad\Longleftrightarrow\quad \text{states built from }\Omega_2,
\\
\text{Group III:} \quad & N_{\rm tri}\neq 0
\quad\Longleftrightarrow\quad \text{states containing }\Omega_3
\text{ and, for longer chains, higher }\Omega_m.
\end{align}
Thus $N_{\rm adj}$ and $N_{\rm tri}$ resolve the first three layers of the
hierarchy, while a complete classification of all dark states requires the full
set of block correlators $N_\ell$.

\section{Adjacent excitations as an entanglement witness in the EAST dark manifold}
\label{app:Nadj_witness}

In the standard Dicke model, the collective decay dynamics starting from the fully inverted state remains separable at all times meaning it does not exhibit genuine many-body entanglement. In our constrained model, by contrast, the structure of the dark manifold allows for a simple witness of genuine many-body entanglement. Here we show that any separable dark state of the EAST-type collective jump operator must have vanishing adjacent-density correlator. Consequently, any dark state with nonzero adjacent-density correlator is necessarily entangled.

We consider the periodic EAST jump operator
\begin{equation}
F=\sum_{j=1}^N n_{j-1}\sigma_j^-,
\qquad
n_j=\frac{\mathbb{I}+\sigma_j^z}{2},
\label{eq:F_East_appendix}
\end{equation}
with site indices understood modulo $N$, and define the adjacent-density operator
\begin{equation}
N_{\mathrm{adj}}=\sum_{j=1}^N n_j n_{j+1}.
\label{eq:Nadj_def_appendix}
\end{equation}
We will prove the following statement.

\paragraph*{Proposition.}
Let $\rho_{\mathrm{sep}}$ be a separable density matrix satisfying
\begin{equation}
\text{Tr}\!\left(\rho_{\mathrm{sep}} F^\dagger F\right)=0.
\label{eq:dark_condition_sep}
\end{equation}
Then
\begin{equation}
\text{Tr}\!\left(\rho_{\mathrm{sep}} N_{\mathrm{adj}}\right)=0.
\label{eq:sep_implies_zero_adj}
\end{equation}
Equivalently, any dark state $\rho$ with
\begin{equation}
\text{Tr}\!\left(\rho N_{\mathrm{adj}}\right)>0
\label{eq:witness_condition}
\end{equation}
cannot be separable and is therefore entangled.

\paragraph*{Proof.}
We begin with a pure product state
\begin{equation}
\ket{\phi}=\bigotimes_{j=1}^N \ket{\phi_j}.
\label{eq:product_state_def}
\end{equation}
For each site, define the local expectation values
\begin{equation}
p_j:=\bra{\phi}n_j\ket{\phi},
\qquad
s_j:=\bra{\phi}\sigma_j^-\ket{\phi}.
\label{eq:local_p_s_def}
\end{equation}
Since $\ket{\phi}$ factorizes across sites, expectation values of operators acting on different sites factorize as well. Writing
\begin{equation}
F^\dagger F=\sum_{i,j}\sigma_i^+ n_{i-1} n_{j-1}\sigma_j^-,
\label{eq:FdagF_expand}
\end{equation}
we separate diagonal and off-diagonal contributions in the site indices. For $i\neq j$, the expectation value factorizes completely,
\begin{equation}
\bra{\phi}\sigma_i^+ n_{i-1} n_{j-1}\sigma_j^-\ket{\phi}
=
p_{i-1} p_{j-1} s_i^\ast s_j,
\qquad i\neq j.
\label{eq:offdiag_factorization}
\end{equation}
For $i=j$, one uses $n_j \sigma_j^-=\sigma_j^-$ and $\sigma_j^+ n_j=\sigma_j^+$ to obtain
\begin{equation}
\bra{\phi}\sigma_j^+ n_{j-1} n_{j-1}\sigma_j^-\ket{\phi}
=
\bra{\phi}n_{j-1} n_j\ket{\phi}
=
p_{j-1}p_j.
\label{eq:diag_factorization}
\end{equation}
Combining both pieces gives
\begin{align}
\bra{\phi}F^\dagger F\ket{\phi}
&=
\sum_{i\neq j} p_{i-1}p_{j-1}s_i^\ast s_j
+
\sum_j p_{j-1}p_j
\nonumber\\
&=
\left|\sum_j p_{j-1}s_j\right|^2
+
\sum_j p_{j-1}\bigl(p_j-|s_j|^2\bigr).
\label{eq:FdagF_product_identity}
\end{align}
For a pure single-qubit state
\begin{equation}
\ket{\phi_j}=\alpha_j\ket{\uparrow}+\beta_j\ket{\downarrow},
\label{eq:single_qubit_param}
\end{equation}
one has
\begin{equation}
p_j=|\alpha_j|^2,
\qquad
s_j=\beta_j^\ast \alpha_j,
\qquad
|s_j|^2=p_j(1-p_j).
\label{eq:single_site_identity}
\end{equation}
Therefore
\begin{equation}
p_j-|s_j|^2=p_j^2,
\label{eq:p_minus_s_identity}
\end{equation}
and Eq.~\eqref{eq:FdagF_product_identity} simplifies to
\begin{equation}
\boxed{
\bra{\phi}F^\dagger F\ket{\phi}
=
\left|\sum_j p_{j-1}s_j\right|^2
+
\sum_j p_{j-1}p_j^2.
}
\label{eq:FdagF_positive_decomp}
\end{equation}
Both terms on the right-hand side are manifestly nonnegative.

Now suppose that the product state $\ket{\phi}$ is dark, i.e.
\begin{equation}
\bra{\phi}F^\dagger F\ket{\phi}=0.
\label{eq:pure_dark_condition}
\end{equation}
Since Eq.~\eqref{eq:FdagF_positive_decomp} is a sum of nonnegative terms, each term must vanish separately. In particular,
\begin{equation}
\sum_j p_{j-1}p_j^2=0.
\label{eq:sum_vanish_condition}
\end{equation}
Every summand is nonnegative, so
\begin{equation}
p_{j-1}p_j^2=0
\qquad
\text{for all } j.
\label{eq:termwise_zero}
\end{equation}
Hence $p_{j-1}p_j=0$ for every bond, and therefore
\begin{equation}
\bra{\phi}N_{\mathrm{adj}}\ket{\phi}
=
\sum_j \bra{\phi}n_j n_{j+1}\ket{\phi}
=
\sum_j p_j p_{j+1}
=0.
\label{eq:pure_product_zero_adj}
\end{equation}
We have thus shown that every pure product dark state has vanishing adjacent-density correlator.

We now extend the argument to a general separable mixed state. Let
\begin{equation}
\rho_{\mathrm{sep}}=\sum_\lambda w_\lambda \ket{\phi_\lambda}\bra{\phi_\lambda},
\qquad
w_\lambda\ge 0,
\qquad
\sum_\lambda w_\lambda=1,
\label{eq:sep_decomp}
\end{equation}
be a separable decomposition into pure product states, and assume that $\rho_{\mathrm{sep}}$ is dark in the sense of Eq.~\eqref{eq:dark_condition_sep}. Since $F^\dagger F$ is positive semidefinite,
\begin{equation}
0=\text{Tr}\!\left(\rho_{\mathrm{sep}}F^\dagger F\right)
=
\sum_\lambda w_\lambda
\bra{\phi_\lambda}F^\dagger F\ket{\phi_\lambda},
\label{eq:mixed_dark_psd}
\end{equation}
where every term in the sum is nonnegative. It follows that
\begin{equation}
\bra{\phi_\lambda}F^\dagger F\ket{\phi_\lambda}=0
\qquad
\text{for every } \lambda \text{ with } w_\lambda>0.
\label{eq:each_component_dark}
\end{equation}
Each product component is therefore itself dark, and by Eq.~\eqref{eq:pure_product_zero_adj} each satisfies
\begin{equation}
\bra{\phi_\lambda}N_{\mathrm{adj}}\ket{\phi_\lambda}=0.
\label{eq:each_component_zero_adj}
\end{equation}
Consequently,
\begin{equation}
\text{Tr}\!\left(\rho_{\mathrm{sep}}N_{\mathrm{adj}}\right)
=
\sum_\lambda w_\lambda
\bra{\phi_\lambda}N_{\mathrm{adj}}\ket{\phi_\lambda}
=0.
\label{eq:mixed_zero_adj}
\end{equation}
This proves Eq.~\eqref{eq:sep_implies_zero_adj}.

We conclude that within the dark manifold of the EAST-type collective dissipator, the observable $N_{\mathrm{adj}}$ is an entanglement witness:
\begin{equation}
\text{Tr}\!\left(\rho N_{\mathrm{adj}}\right)>0
\quad\Longrightarrow\quad
\rho \text{ is entangled.}
\label{eq:Nadj_witness_final}
\end{equation}

Two points are worth emphasizing. First, the witness is specific to the dark manifold. A generic separable state can of course have $\langle N_{\mathrm{adj}}\rangle>0$; what is special here is the combination of separability and the dark-state constraint. Second, the witness is experimentally simple: it only requires measuring adjacent excitation correlations, rather than reconstructing the full density matrix or evaluating a more elaborate multipartite entanglement measure.

\section{Justification of the rotating wave approximation}

\begin{figure}[ht]
    \centering
    \includegraphics[width=0.8\textwidth]{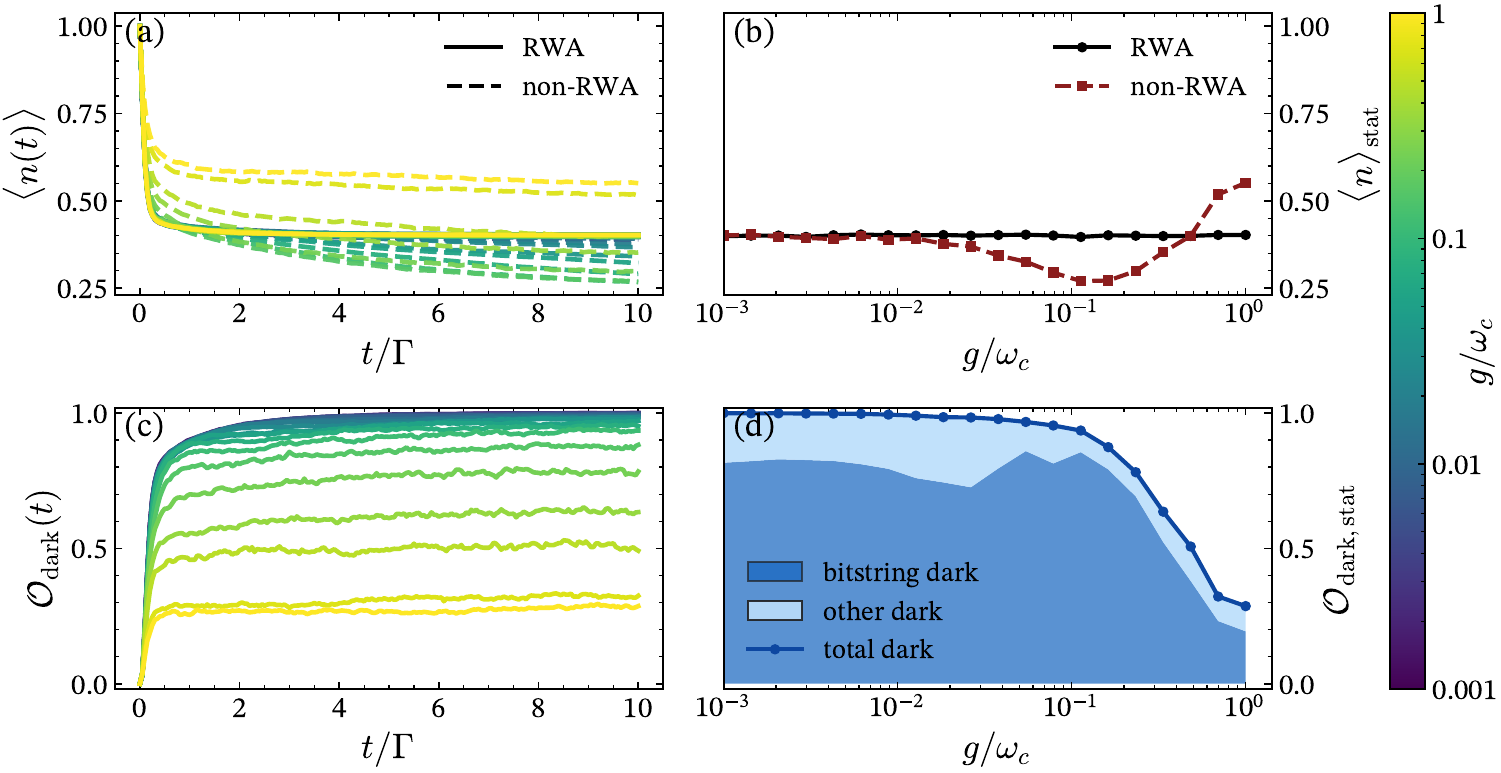}
\caption{\textbf{Integrated-out cavity dynamics: RWA vs non-RWA.}
(a) Time traces of the mean spin excitation $\langle n(t)\rangle=\frac{1}{N}\sum_j\langle n_j(t)\rangle$ for RWA (solid) and non-RWA (dashed), with color encoding the sweep parameter $g/\omega_c$.
(b) Stationary value $\langle n\rangle_{\mathrm{stat}}$ (average over the final 10\% of the simulated time window) versus $g/\omega_c$.
(c) Time dependence of the non-RWA overlap with the RWA dark-state manifold.
(d) Final dark-manifold overlap versus $g/\omega_c$.
Across the sweep, the RWA/non-RWA discrepancy in panel (a) increases from $0.00504$ at $g/\omega_c=0.001$ to $0.167$ at $g/\omega_c=1$; the non-RWA stationary excitation increases from $0.402$ to $0.554$, while the final dark-manifold overlap decreases from $0.999$ to $0.275$.
    Parameters: $N=8$, constraint = EAST, $\Delta=0$, $\kappa=40$, sweep mode = \texttt{fixed\_g}, trajectories = $800$.}
    \label{fig:rwa_comparison}
    
\end{figure}

An important assumption in the main text is the application of the rotating wave approximation (RWA). To analyse the validity of this step we will compare the dynamics in the bad-cavity limit with RWA and without RWA. In Fig. \ref{fig:rwa_comparison}

RWA is an excellent approximation at weak coupling. In Fig.~\ref{fig:rwa_comparison}(a), the RWA (solid) and non-RWA (dashed) time traces of the mean excitation $\langle n(t)\rangle$ are visually indistinguishable for $g/\omega_c \lesssim 10^{-2}$ and relax to the same plateau. Figure~\ref{fig:rwa_comparison}(b) confirms this quantitatively: the stationary value $\langle n\rangle_{\mathrm{stat}}$ agrees between RWA and non-RWA in this perturbative regime.

Virtual (counter-rotating) processes modify the stationary excitation outside the perturbative limit. For intermediate couplings, $10^{-2}\lesssim g/\omega_c \lesssim 10^{-1}$, the non-RWA dynamics relaxes to a smaller $\langle n\rangle_{\mathrm{stat}}$ than in RWA, consistent with additional off-resonant channels that allow the system to escape dark states higher in the decay ladder and relax further. For $g/\omega_c \sim 1$, these virtual fluctuations become strong and instead increase $\langle n\rangle_{\mathrm{stat}}$, producing the upturn in Fig.~\ref{fig:rwa_comparison}(b).

The non-RWA trajectories initially remain confined to the RWA dark manifold. Figure~\ref{fig:rwa_comparison}(c) shows that the overlap $\mathcal{O}_{\mathrm{dark}}(t)$ quickly approaches unity for $g/\omega_c \lesssim 10^{-1}$, indicating that the same fragmented dark subspace organizes the early and intermediate-time relaxation.

At stronger coupling, virtual fluctuations reduce and reshape the dark-manifold weight. For $g/\omega_c \gtrsim 10^{-1}$, $\mathcal{O}_{\mathrm{dark}}(t)$ saturates at progressively smaller values, and the stationary overlap $\mathcal{O}_{\mathrm{dark,stat}}$ drops sharply as $g/\omega_c \to 1$ [Fig.~\ref{fig:rwa_comparison}(d)]. Moreover, before the total overlap collapses, the dark-sector composition shifts: weight is funneled toward bitstring dark states at the expense of more correlated dark states. Overall, Fig.~\ref{fig:rwa_comparison} shows that RWA remains reliable for $g/\omega_c \lesssim 10^{-2}$ (and qualitatively accurate up to $\sim 10^{-1}$), while non-RWA virtual processes control both $\langle n\rangle_{\mathrm{stat}}$ and the late-time dark-state structure at larger $g/\omega_c$.

\section{Discrete truncated Wigner approximation (DTWA) calculations}
To study the dynamics of large systems beyond the regime accessible to exact master-equation simulations, we employ a phase-space approach based on the truncated Wigner approximation. The method combines a discrete phase-space representation for the spin degrees of freedom~\cite{Wootters1987Wigner} with a continuous-variable Wigner representation for the cavity field~\cite{Polkovnikov2010PhaseSpace}, allowing for an efficient treatment of dissipative many-body dynamics while retaining leading-order quantum fluctuations.

In this representation, quantum operators are mapped onto classical phase-space variables whose dynamics follow semiclassical equations of motion. Quantum fluctuations are incorporated through stochastic sampling of the initial Wigner distribution and through noise associated with cavity dissipation. The approximation consists in truncating higher-order derivatives in the phase-space evolution equation, thereby neglecting higher-order quantum correlations. Expectation values are then reconstructed from ensemble averages over classical trajectories.

For our calculations, we consider the Hamiltonian (1) from the main text, 
\begin{align}
    \hat{H} = \hbar \Delta \hat{a}^\dagger \hat{a} + \hbar g ( \hat{F} \hat{a}^\dagger + \hat{F}^\dagger \hat{a}) \, ,
\end{align}
where $\hat{a}$ ($\hat{a}^\dagger$) describes the annihilation (creation) operator of the cavity photon, $g$ is the atom-cavity coupling, and $\Delta$ is the detuning between the cavity frequency and the atomic transition frequency. The operator $\hat{F}$ accounts for the kinetically constrained coupling
\begin{align}
    \hat{F} = \sum_j \hat{P}_j \hat{\sigma}_j^- \, ,
\end{align} 
with $\hat{\sigma}_j^-$ being the spin-1/2 lowering operator for atom $j$ and $\hat{P}_j$ being the projector enforcing the kinetic constraint. For the specific models we consider in the main text, the projector takes the form
\begin{align}
    \hat{P}_j = \begin{cases} \hat{I}  \, , &\quad \alpha = \beta = \gamma = 0 \\
        \alpha \hat{n}_{j-1} + \beta \hat{n}_{j+1} + \gamma \hat{n}_{j-1} \hat{n}_{j+1} \, , &\quad \mathrm{else}
    \end{cases} \, . 
    \label{eq:supp_constraint_projector}
\end{align}
Here, $\hat{I}$ is the identity operator and $\hat{n} = (\hat{I} + \hat{\sigma}_z)/2$. Different choices $(\alpha,\beta,\gamma)$ correspond to the models considered in the main text according to
\begin{center}
    \begin{tabular}{c|c|c|c}
        {} & $\alpha$ & $\beta$ & $\gamma$ \\
        \hline
        Dicke & 0 & 0 & 0 \\
        \hline
        East & 1 & 0 & 0 \\
        \hline
        AND & 0 & 0 & 1 \\
        \hline
        OR & 1 & 1 & -1
    \end{tabular} \, . 
\end{center}

The full time evolution is given by the master equation
\begin{align}
    \dot{\hat{\rho}} = - \frac{i}{\hbar} [ \hat{H}, \hat{\rho}] + \kappa \left( \hat{a} \hat{\rho} \hat{a}^\dagger - \frac{1}{2} \{ \hat{a}^\dagger \hat{a} , \hat{\rho} \} \right) 
\end{align}
accounting for the cavity loss with rate $\kappa$. 

Following Ref.~\cite{HosseinabadiChelpanovaMarino2025UserFriendlyDTWA} we derive the semiclassical stochastic equations of motion for the classical variables $a$, $s_j^-$ and $s_j^z$ describing the cavity field and spin variables, respectively, which read
\begin{align}
    \dot{a} &= - i \Delta a - i g \sum_j P_j s^-_j - \frac{\kappa}{2} a - \frac{1}{2} \xi(t) \\
    \dot{s}_j^- &= i g a P_j s^z_j - i g s^-_j \left[ \alpha a s_{j+1}^+ + \beta a s_{j-1}^+ + \gamma a \left( s_{j+1}^+ n_{j+2} + n_{j-2} s_{j-1}^+\right) + \mathrm{c.c.}\right]  \\
    \dot{s}_j^z &= - 2 i g P_j (a s_j^+ - a^\dagger s_j^-) \, . 
\end{align} 
Here, $s_j^\pm = s_j^x \pm i s_j^y$, $n_j = (1 + s_j^z)/2$, and $P_j$ is obtained by replacing operators with corresponding phase-space variables in Eq.~(\ref{eq:supp_constraint_projector}). The noise process $\xi$ arises from the Wigner representation of cavity loss and corresponds to white noise in the Ito convention. It thus satisfies $\langle \langle \xi(t) \rangle \rangle = 0$ and $\langle \langle \xi(t) \xi^*(t') \rangle \rangle = 2 \kappa \delta (t - t')$, with $\langle \langle \cdots \rangle \rangle$ denoting the average over different noise realizations. Unless stated otherwise, periodic boundary conditions are assumed.

The initial Wigner distribution is obtained by sampling different initial configurations. The discrete Wigner representation of a spin-polarized state samples transverse spin components from the discrete set $s_j^{x,y} = \pm 1$, while $s_j^z$ is fixed by the initial polarization. In our specific case, we consider an initial state with all spins pointing up and the cavity field being in a coherent state with parameter $\alpha_0 = e^{i \phi} \sqrt{\bar{n}_c}$, where $\phi$ denotes the phase of the coherent state and $\bar{n}_c$ its mean occupation number. This leads to the following initial conditions
\begin{align}
    a &= \alpha_0 + \delta a \, , \\
    s_j^x &= \pm 1 \, , \\
    s_j^y &= \pm 1 \, , \\
    s_j^z & = 1 \, ,
\end{align}
where $\delta a = (\eta_R + i \eta_I)/2$ are random variables with $\eta_R, \eta_I \sim \mathcal{N}(0,1)$. The fluctuations $\delta a$ in the cavity field are necessary for triggering the initial coupling to the cavity mode. The transverse spin components are sampled independently with equal probability, which reduces sampling noise and improves numerical convergence~\cite{MinkPetrosyanFleischhauer2022HybridDTWA}.

Observables are calculated by averaging over different trajectories according to
\begin{align}
    \langle \hat{a}^\dagger \hat{a} \rangle &= \frac{1}{n_\mathrm{traj}} \sum_{m=1}^{n_\mathrm{traj}}  \vert a_m \vert^2 - \frac{1}{2} \\
    \langle \hat{\sigma}_j^\alpha \rangle &= \frac{1}{n_\mathrm{traj}} \sum_{m=1}^{n_\mathrm{traj}} s_{j,m}^\alpha  \\
    \frac{1}{2} \left\langle \left\{ \hat{\sigma}_j^\alpha(t), \hat{\sigma}_l^\beta(t) \right\} \right\rangle &=  \frac{1}{n_\mathrm{traj}} \sum_{m=1}^{n_\mathrm{traj}}  s_{j,m}^\alpha(t) s_{l,m}^\beta(t)  \, .
\end{align}
Each trajectory includes independent sampling of both the initial Wigner distribution and the stochastic noise process. The additional term of $-1/2$ in the photon number operator comes from the fact that the Wigner representation corresponds to symmetric operator ordering.

\section{Comparison between DTWA and quantum-jump simulations}

\begin{figure}[h]
    \centering
    \includegraphics[width=\linewidth]{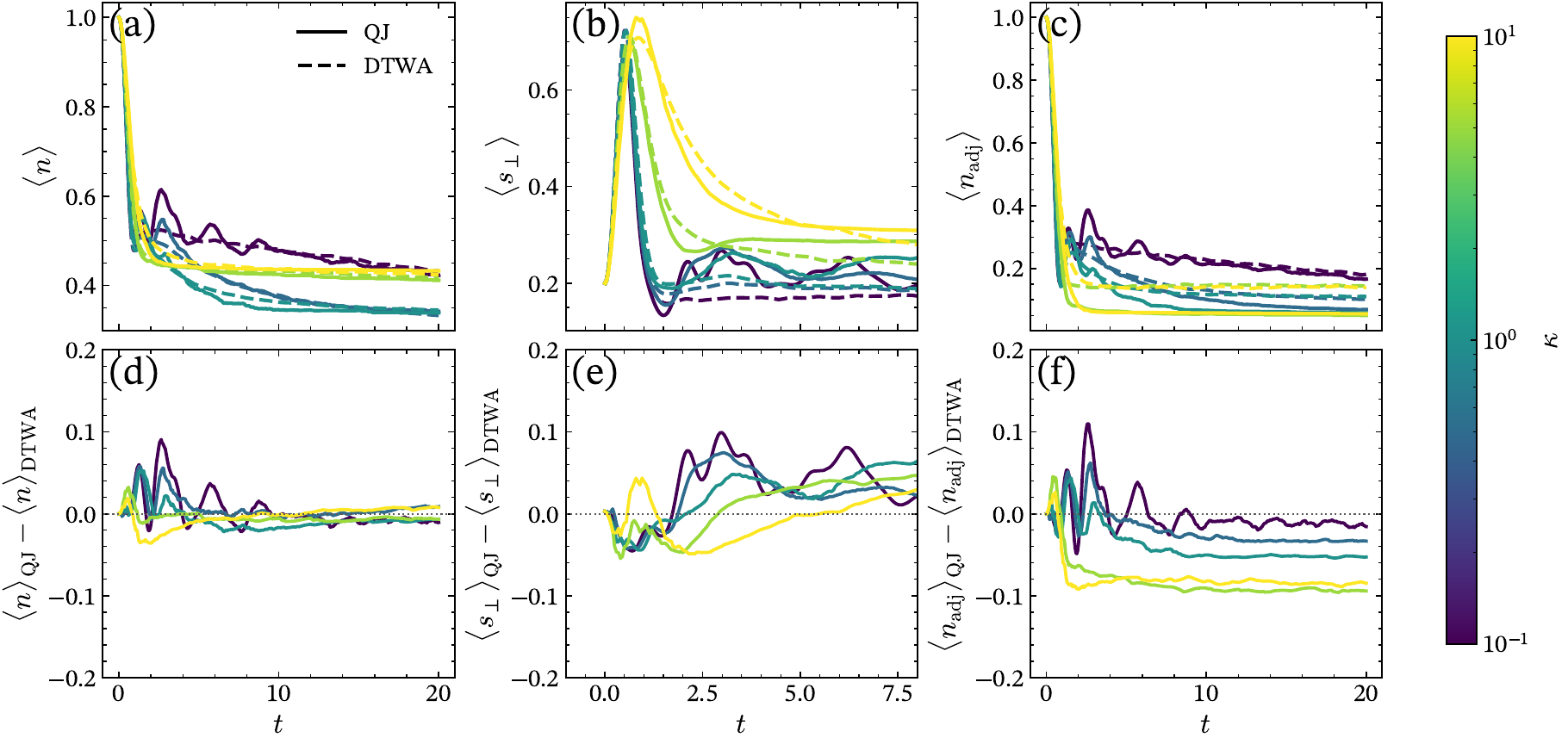}
    \caption{\textbf{Benchmark of DTWA against quantum-jump simulations.}
(a) Mean excitation density $\langle n(t)\rangle=\frac{1}{N}\sum_j\langle n_j(t)\rangle$,
(b) collective transverse coherence $\langle S_\perp^2(t)\rangle\equiv\langle S_x^2+S_y^2\rangle$ (superradiant burst),
and (c) nearest-neighbor correlator $\langle N_{\mathrm{adj}}(t)\rangle\equiv \sum_j\langle n_j n_{j+1}\rangle$
for several dissipation rates $\kappa$ (color scale).
Solid lines show numerically exact quantum-jump (QJ) results, while dashed lines show the dynamical truncated Wigner approximation (DTWA).
(d--f) Relative deviations $(\langle O\rangle_{\mathrm{QJ}}-\langle O\rangle_{\mathrm{DTWA}})/\langle O\rangle_{\mathrm{DTWA}}$ for the corresponding observables.
DTWA reproduces $\langle n(t)\rangle$ quantitatively at all times and captures the height and timing of the superradiant peak in $\langle S_\perp^2(t)\rangle$ with high accuracy; the residual discrepancies are small and typically decrease at late times and for larger $\kappa$.
By contrast, DTWA can fail for higher-order correlations in the stationary state: the long-time value of $\langle N_{\mathrm{adj}}\rangle$ shows pronounced, order-unity deviations at large $\kappa$, consistent with the importance of quantum coherences within the fragmented dark manifold that are not retained under cumulant truncation.
    Parameters: EAST constraint on a periodic chain of $N$ spins, rotating-wave approximation, $\Delta=0$, initial state $|e\rangle^{\otimes N}$, and $\kappa$ as indicated.}
    \label{fig:dtwaVQquantumJump}
\end{figure}

To evaluate how superradiant emission scales in large systems, we employ the dynamical truncated Wigner approximation (DTWA). In DTWA, higher-order cumulants are neglected by factorizing them into lower-order correlators, yielding a tractable semiclassical description of the dynamics. In Fig.~\ref{fig:dtwaVQquantumJump}, we benchmark DTWA against numerically exact quantum-jump simulations.

Figure~\ref{fig:dtwaVQquantumJump} benchmarks the dynamical truncated Wigner approximation (DTWA, dashed) against numerically exact quantum-jump simulations (QJ, solid) for several dissipation rates $\kappa$ (color scale).  DTWA accurately reproduces the early-time collective emission dynamics that underlies the superradiant burst: in particular, the transverse coherence $\langle S_\perp^2\rangle$ in panel~(b) shows a superradiant peak whose height and timing are in close agreement with QJ across the explored $\kappa$ range.  The mean excitation density $\langle n\rangle$ in panel~(a) is likewise captured very well at all times, indicating that DTWA provides a reliable description of the one-body relaxation dynamics relevant for extracting superradiant growth and its scaling with system size.

The bottom row quantifies the relative deviations $(\langle O\rangle_{\rm QJ}-\langle O\rangle_{\rm DTWA})/\langle O\rangle_{\rm DTWA}$ for each observable.  For $\langle n\rangle$ and $\langle S_\perp^2\rangle$ [panels~(d,e)], the discrepancies remain small and typically diminish as the dynamics proceeds; agreement also improves with increasing $\kappa$.  By contrast, DTWA can fail for higher-order correlators that probe the correlated dark manifold: panel~(c) shows that the nearest-neighbor correlator $\langle N_{\rm adj}\rangle \equiv \sum_j\langle n_j n_{j+1}\rangle$ exhibits pronounced long-time deviations, which become most severe at large $\kappa$, reaching order-unity relative errors in the stationary state [panel~(f)].  This breakdown is expected because $\langle N_{\rm adj}\rangle$ is controlled by quantum coherences and interference within the fragmented dark-state manifold, which are not faithfully captured once higher-order cumulants are truncated.

Overall, DTWA is a useful and computationally efficient tool for large-system scaling analyses of the early-time superradiant burst--most importantly the peak in $\langle S_\perp^2\rangle$ and the accompanying decay of $\langle n\rangle$--while stationary-state conclusions for multi-body correlators such as $\langle N_{\rm adj}\rangle$ require fully quantum methods.

\section{Experimental implementation using Rydberg atoms}

Here, we consider a possible experimental implementation using Rydberg atoms coupled to an optical cavity based on Ref.~\cite{ValenciaTortoraEtAl2024}. We start by considering an ensemble of three-level atoms with internal states $\ket{g}$, $\ket{e}$, and $\ket{r}$ representing the ground state, the intermediate state, and the Rydberg state, respectively. Their energies are, respectively, $\omega_g = 0$, $\omega_e$, and $\omega_r$. The intermediate state has a linewidth $\gamma_e$, while the Rydberg state is assumed to be long-lived with $\gamma_r \approx 0$. The $\ket{g} - \ket{e}$ transition is coupled to an optical cavity with resonance frequency $\omega_c$ and linewidth $\kappa$ and the $\ket{e}-\ket{r}$ transition is coupled via a classical laser field with time and position-dependent Rabi frequency $\Omega_j(t) = \Omega f_j(t)$, where the time-dependence will be specified below. The Rydberg states interact via a van der Waals interaction $V_{jl} = C_6 / \vert r_j - r_l \vert^6$ with atomic positions $r_j$. The master equation for this system reads
\begin{align}
    \partial_t \rho = - \frac{i}{\hbar} \left[ H, \rho \right] + \kappa \left( a \rho a^\dagger - \frac{1}{2}\left\{a^\dagger a, \rho \right\} \right) + \gamma_e \sum_j \left( \sigma_j^{ge} \rho \sigma_j^{eg} - \frac{1}{2} \left\{ \sigma_j^{eg} \sigma_j^{ge} , \rho\right\}\right)
\end{align}
where $\sigma_j^{\alpha \beta} = \vert{\alpha}\rangle \langle{\beta}\vert$ and the Hamiltonian in an appropriate rotating frame
\begin{align}
    \frac{H}{\hbar} = \sum_j \left( - \Delta \sigma_j^{ee} - (\Delta + \delta) \sigma_j^{rr} \right) + g \sum_j \left( a \sigma_j^{eg} + a^\dagger \sigma_j^{ge} \right) + \Omega \sum_j \left( \sigma_j^{er} f_j^*(t) + f(t) \sigma_j^{re} \right) + \frac{1}{2} \sum_{j \neq l} V_{jl} \sigma_j^{rr} \sigma_l^{rr} \, .
\end{align}
In this Hamiltonian, the first term describes the energies of the atomic levels in the rotating frame with $\Delta = \omega_c - \omega_e$ and $\delta$ an additional two-photon detuning, the second term describes the coupling of the atoms to the cavity, the third term the coupling by the classical laser field, and the last term describes the Rydberg interaction between the atoms.

Next, we adiabatically eliminate the intermediate state $\ket{e}$ in the limit of large intermediate-state detuning $\Delta$. This leads to the effective low-energy Hamiltonian
\begin{align}
    \frac{H}{\hbar} = - \sum_j \delta_{\mathrm{eff},j}(t) \sigma_j^{rr} + g_\mathrm{eff} \sum_j \left( f_j(t) a \sigma_j^{rg} + f_j^*(t) a^\dagger \sigma_j^{gr} \right) + \frac{1}{2} \sum_{j \neq l} V_{jl} \sigma_j^{rr} \sigma_l^{rr}
\end{align}
where $\delta_{\mathrm{eff}, j}(t) = \delta + g^2 / \Delta a^\dagger a - \Omega^2 \vert f_j(t) \vert^2 / \Delta$ is a time- and position-dependent effective detuning due to AC Stark shifts coming from the classical drive as well as from the cavity field. The AC Stark shift coming from the latter is photon number dependent. The effective Raman coupling of the ground state to the Rydberg state is given by $g_\mathrm{eff} = g \Omega /\Delta$. Furthermore, after the adiabatic elimination, the effective decay rate from the Rydberg state will be $\gamma_\mathrm{eff} = \gamma_e (\Omega / \Delta)^2$. 

\subsubsection{Implementing the EAST constraint}
We first focus on implementing the kinetic constraint for the EAST model. To this end, we first position the atoms in a staggered configuration such that the distance between an odd and an even site is given by $r_1$ and the distance between an even site and the next odd site to the right by $r_2$. We further denote $V_1 = C_6/r_1^6$ and $V_2 = C_6 / r_2^6$. Thus, assuming we are at an odd site $j$ and the atom to the left is in the Rydberg state, the Rydberg state at site $j$ experiences an interaction-induced shift by $V_2$. 

The EAST constraint can then be implemented by taking the time dependence of the classical drive to be
\begin{align}
    f_j(t) = e^{- i t V_{j-1,j}} + e^{ - i t (V_1 + V_2)} \, .
\end{align}

This can be realized by having two laser beams with the same Rabi frequency $\Omega$ but one detuned by $V_{j-1,j}$ and the other by $V_1 + V_2$. While the first laser will be in resonance only when the atom to the left is in the Rydberg state, the second laser, detuned by $V_1 + V_2$ will make the transition resonant if both neighboring atoms are in the Rydberg state. We will see below that for the XOR and AND model, only one laser will be necessary and the constraints can also be implemented by means of local light shifts.

Next, we go into an interaction picture with respect to the Hamiltonian $H_V = \hbar\sum_j V_{j, j+1} \sigma_j^{rr} \sigma_{j+1}^{rr}$, that is 
\begin{align}
    H_\mathrm{int} = U^\dagger(t) H U(t) - H_V \, , \qquad U(t) = \exp( - i t H_V / \hbar) \,. 
\end{align}
For the transformed Hamiltonian, we only need the time evolution of the operator $\sigma_l^{re}$ since all other parts will commute with $U(t)$ such that
\begin{align}
    U^\dagger(t) \sigma_l^{re} U(t) &= \exp \left[i t \left( V_{l, l+1} \sigma_l^{rr} \sigma_{l+1}^{rr}  + V_{l-1,l} \sigma_{l-1}^{rr} \sigma_l^{rr}\right)/\hbar \right] \sigma_l^{re} \exp\left[ - i t \left( V_{l,l+1} \sigma_l^{rr} \sigma_{l+1}^{rr} + V_{l-1,l} \sigma_{l-1}^{rr} \sigma_l^{rr} \right)/\hbar \right] \, \nonumber \\
    &= \exp( O_l \sigma_l^{rr}) \sigma_l^{re} \exp( O_l \sigma_l^{rr}) 
\end{align}
with $O_l = (V_{l,l+1} \sigma_{l+1}^{rr} + V_{l-1,l} \sigma_{l-1}^{rr})/\hbar$. Using $[\sigma_l^{rr}, \sigma_j^{re}] = \delta_{jl} \sigma_l^{re}$ we also have $[O_l\sigma_{l}^{rr}, \sigma_l^{re}] = \sigma_l^{re} O_l$ and further $[O_l\sigma_{l}^{rr}, \sigma_l^{re}]_n = \sigma_l^{re} O_l^n$, where $[A,B]_n$ is a nested commutator. Then, we can calculate
\begin{align}
    \exp( O_l \sigma_l^{rr}) \sigma_l^{re} \exp( O_l \sigma_l^{rr}) &= \sum_n \frac{O_l^n}{n!} \sigma_l^{re} \nonumber \\
    &= \exp(O_l) \sigma_l^{re} \nonumber \\
    &= \exp\left( i t V_{l-1,l} \sigma_{l-1}^{rr}  / \hbar\right)\sigma_l^{re} \exp\left( i t V_{l,l+1} \sigma_{l+1}^{rr} / \hbar\right) \nonumber \\
    &= \left[ 1 + \sum_{n=1} \frac{( i t V_{l-1,l}/\hbar)^n}{n!} \sigma_{l-1}^{rr} \right] \sigma_l^{re} \left[1 + \sum_{n=1} \frac{(i t V_{l,l+1}/\hbar)^n}{n!} \sigma_{l+1}^{rr} \right] \nonumber \\
    &= \left[ 1 + \left( e^{i t V_{l-1,l}/\hbar} - 1\right)\sigma_{l-1}^{rr} \right] \sigma_l^{re} \left[1 + \left( e^{i t V_{l,l+1}/\hbar} - 1\right)\sigma_{l+1}^{rr} \right] \nonumber \\
    &= \left[ P^g_{l-1} + e^{ i t V_{l-1,l}/\hbar} \sigma_{l-1}^{rr} \right] \sigma_l^{re} \left[ P^g_{l+1} + e^{ i t V_{l+1,l}/\hbar} \sigma_{l+1}^{rr} \right]
\end{align}
with the $P^g_l = 1 - \sigma_l^{rr}$ being the projector on the ground state at site $l$. The Hamiltonian then takes the form
\begin{align}
    \frac{H_\mathrm{int}}{\hbar} &= - \sum_j \delta_{\mathrm{eff},j} \sigma_j^{rr} + \frac{1}{2} \sum_{ j \neq l} V_{jl} \sigma_j^{rr} \sigma_j^{rr} - \sum_j V_{j, j+1} \sigma_j^{rr} \sigma_{j+1}^{rr} \nonumber \\
    & \qquad + g_\mathrm{eff} \bigg\lbrace f_1(t) \sigma_1^{re} \left( P_2 + e^{ i V_1 t /\hbar} \sigma_2^{rr} \right) + \mathrm{h.c.} \nonumber \\
    & \qquad + \sum_{j=2}^{N-1} f_j(t) \left( P_{j-1} + e^{ i V_{j-1,j} t /\hbar} \sigma_{j-1}^{rr} \right) \sigma_j^{re} \left( P_{j+1} + e^{ i V_{j,j+1}t /\hbar} \sigma_{j+1}^{rr} \right) + \mathrm{h.c.} \nonumber \\
    & \qquad + f_N(t) \left( P_{N-1} + e^{i V_{N-1,N}t /\hbar}\right) \sigma_N^{re} + \mathrm{h.c.}\bigg\rbrace \, . 
\end{align}
We have
\begin{align}
    V_{j,j+1} = \begin{cases}
        V_1, \quad  j \, \, \mathrm{odd} \\
        V_2, \quad j \, \, \mathrm{even} \, 
    \end{cases}
\end{align}
and we take only the resonant terms such that for $V_1, V_2, \vert V_1 - V_2 \vert \gg g_\mathrm{eff}, \gamma_\mathrm{eff}, \kappa$, we get, for example, for the odd sites
\begin{align}
    \left( e^{- i t V_2} + e^{- i t (V_1 + V_2)} \right) \left( P_{j-1} + e^{ i V_2 t} \sigma_{j-1}^{rr} \right) \sigma_j^{re} \left( P_{j+1} + e^{i V_2 t} \sigma_{j+1}^{rr} \right) &\approx \sigma_{j-1}^{rr} \sigma_j^{re} P_{j+1} + \sigma_{j-1}^{rr} \sigma_j^{re} \sigma_{j+1}^{rr} \nonumber \\
    & = \sigma_{j-1}^{rr} \sigma_j^{re},
\end{align}
and similarly for the even sites, which implements the kinetic constraint of the EAST model. 

The Hamiltonian after the adiabatic elimination of the intermediate state also contains a time-dependent AC Stark shift 
\begin{align}
    \frac{\Omega^2 \vert f(t) \vert^2}{ \Delta} = \frac{\Omega^2}{\Delta} = 2\frac{\Omega^2}{\Delta} + \frac{\Omega^2}{\Delta} \cos( (V_1 + V_2 + V_{j-1,j})t) \, . 
\end{align}
In the same limit, where we can neglect the off-resonant terms to obtain the kinetic constraint for the EAST model, the time-dependent term of the AC Stark shift will oscillate much faster than the typical atomic time scales such that this averages out and we can neglect it. 

The final Hamiltonian then reads
\begin{align}
    \frac{H_\mathrm{East}}{\hbar} = - \sum_j \left( \delta + \frac{g^2}{\Delta} a^\dagger a - \frac{2 \Omega^2}{\Delta} \right) \sigma_j^{rr} + g_\mathrm{eff} \sum_{j=2}^N \sigma_{j-1}^{rr} \left( a \sigma_j^{rg} + a^\dagger \sigma_j^{gr} \right) + \sum_{j=1}^{N} \sum_{l = j+2}^{N} V_{jl} \sigma_j^{rr} \sigma_l^{rr} \, ,
\end{align}
where the second term directly gives the EAST-constrained two-photon coupling to the cavity and the last term describes the influence of the residual interactions beyond nearest neighbors. 

\subsubsection{Implementing the XOR and AND constraints}
For the XOR and AND constraints, it is possible to find a simpler description using a single coupling laser for the upper transition. There are then two ways how to implement the constraint. In the first approach, we place the atoms without a staggered configuration with a distance $a$ between them and define $V = C_6/a^6$. For the XOR model, the coupling laser is then detuned from two-photon resonance by $V$ such that if on a given site one of the two neighboring atoms is in the Rydberg state, the laser becomes resonant while being off-resonant if neither or both neighboring atoms are in the Rydberg state. In the AND model, the laser is detuned by $2V$ such that it is only resonant when both neighboring atoms are in the Rydberg state. 

The second, and equivalent, approach is by applying light shifts to detune the atoms such that they become resonant in the presence of interaction shifts. For this, we simply apply the detuning
\begin{align}
    \delta = \begin{cases}
        - V, \quad \mathrm{XOR} \\
        - 2 V, \quad \mathrm{AND} \, . 
    \end{cases}
\end{align}
Note, that we always assume that at least the constant detuning due to the AC Stark shift from the coupling laser is canceled as well. 

\subsection{Parameter estimation}

Recently, arrays of Rydberg atoms in an optical cavity have successfully been realized experimentally~\cite{SantisRydbergCavity2026}. The experiment used ${}^{87}$Rb and a two-photon excitation scheme to couple the ground state $\ket{5S_{1/2}}$ to the Rydberg state $\ket{53 S_{1/2}}$ via the intermediate low-lying state $\ket{5 P_{3/2}}$. For this transition, the linewidth is $\gamma_e = 2 \pi \times 6 \, \mathrm{MHz}$. The cavity linewidth in the experiment is given by $\kappa = 2\pi \times 0.85 \, \mathrm{MHz}$ and given their single-atom cooperativity $C \approx 1$, we obtain $g \approx 2\pi \times 2.26 \, \mathrm{MHz}$. 

For the intermediate state detuning we choose $\Delta = 2\pi \times 100 \, \mathrm{MHz}$, and for the Rabi frequency of the upper transition, we take $\Omega = 2\pi \times 10 \, \mathrm{MHz}$, which are both realistic values for such an experiment, such that $\Omega / \Delta = 0.1$. 

The effective decay is then $\gamma_\mathrm{eff} = \gamma_e (\Omega / \Delta)^2 = 2 \pi \times 0.06 \, \mathrm{MHz} \ll \kappa$, and the effective coupling strength is $g_\mathrm{eff} = g (\Omega / \Delta) = 2 \pi \times 0.23 \, \mathrm{MHz} < \kappa$. Note, however, that $g_\mathrm{eff} N \gg \kappa$ even for moderate $N > 20$. In this regime, a treatment of an adiabatically eliminated cavity would no longer be correct. We can, however, either decrease the ratio $\Omega/\Delta$ or increase the cavity linewidth further. The effect of the photon-number dependent light shift can be estimated from $g^2 / \Delta = 2 \pi \times 0.02 \, \mathrm{MHz} \ll g_\mathrm{eff} , \kappa$. Even for a large number of atoms, say $N = 200$, we get a maximum shift of $g^2 / \Delta \langle a^\dagger a \rangle \leq 2 \pi \times 4 \, \mathrm{MHz}$. 

In order for the kinetic constraints to work, we need strong interactions with $V_1, V_2, \vert V_1 - V_2 \vert \gg g_\mathrm{eff}, \kappa, \gamma_\mathrm{eff}, V_\mathrm{NNN}$, with $V_{NNN}$ being the next-nearest neighbor interaction shift. For the $53 S_{1/2}$ Rydberg state of ${}^{87}$Rb, we get $C_6 / \hbar \approx 2 \pi \times 30 \, \mathrm{GHz} \times \mu \mathrm{m}^6$ with the $C_6$ coefficient scaling as $n^{11}$ with the principal quantum number $n$. For realistic distances of $r_1 = 2 \, \mu \mathrm{m}$ and $r_2 = 2.3 \, \mu \mathrm{m}$, we get $V_1 / \hbar = 2 \pi \times 468 \, \mathrm{MHz}$ and $V_2/ \hbar = 2\pi \times 265 \, \mathrm{MHz}$, such that $\vert V_1 -  V_2 \vert / \hbar = 2\pi \times 203 \, \mathrm{MHz} \gg g_\mathrm{eff}, \kappa, \gamma_\mathrm{eff}$. The next-nearest neighbor interaction shift is $V_\mathrm{NNN}/ \hbar = 2 \pi \times 4.76 \, \mathrm{MHz} \ll V_1/\hbar, V_2/\hbar, \vert V_1 - V_2 \vert / \hbar$. It is important to note, however, that the next-nearest neighbor interaction shift is not much smaller than the cavity linewidth or the effective two-photon coupling strength and we will discuss its influence in Sec.~\ref{app:rydberg_interaction}. Increasing the distances to $r_1 = 2.8 \, \mu \mathrm{m}$ and $r_2 = 3 \, \mu \mathrm{m}$ gives $V_\mathrm{NNN} / \hbar = 2 \pi \times 0.79 \, \mathrm{MHz} \lesssim \kappa$ with the caveat that $\vert V_1 - V_2 \vert / \hbar = 2 \pi \times 21 \, \mathrm{MHz}$.

\section{Robustness to experimental imperfections}

\subsection{Next-nearest-neighbor Rydberg interactions}
\label{app:rydberg_interaction}

\begin{figure}
    \centering
    \includegraphics[width=0.8\linewidth]{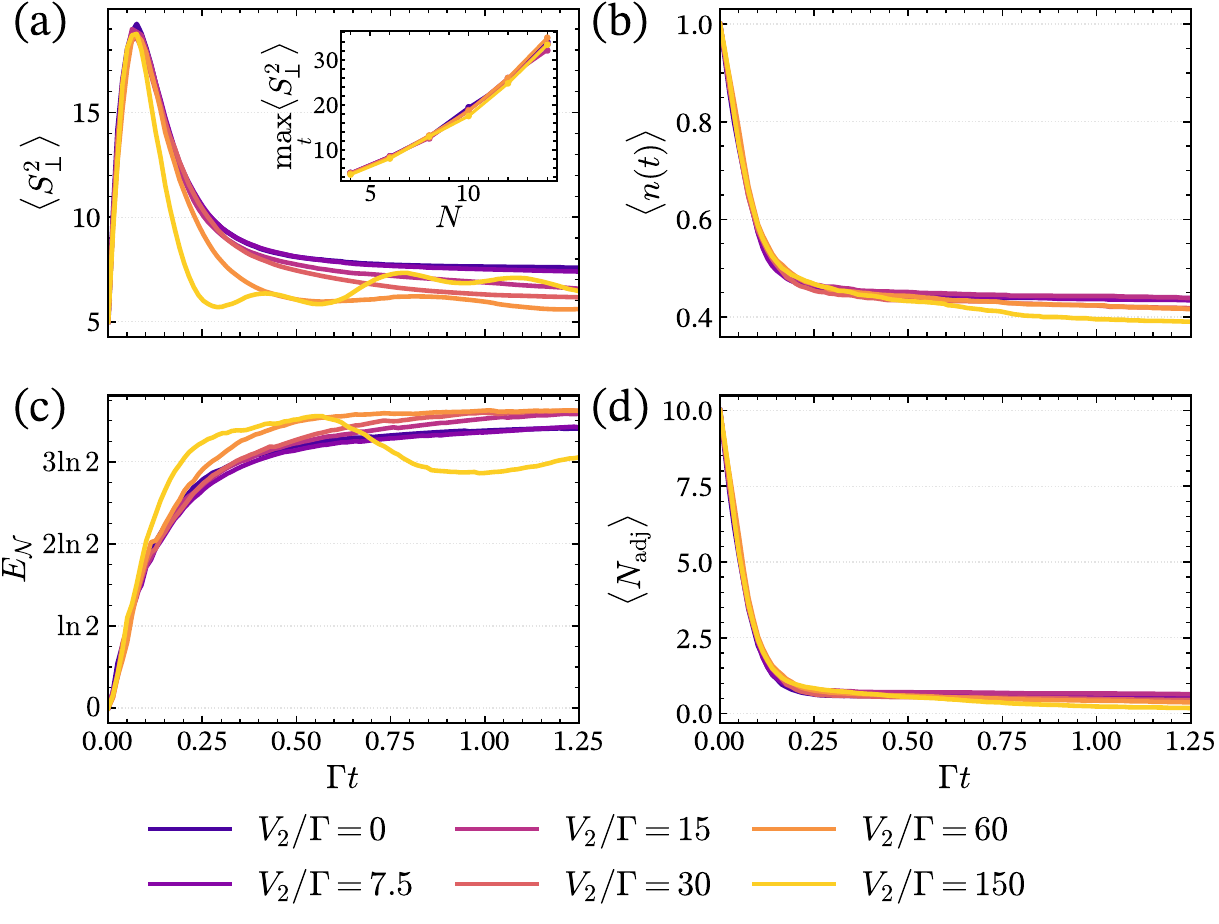}
    \caption{\textbf{Effect of next-nearest-neighbor Rydberg tails on the EAST-model dynamics.}
    Time evolution for a periodic chain with $N=10$ and $\Delta=0$, for several values of the residual tail strength $V_2/\Gamma$ as indicated in the legend.
    (a) Collective transverse coherence $\langle S_\perp^2(t)\rangle\equiv \langle S_x^2+S_y^2\rangle$; the inset shows the system-size scaling of the peak value $\max_t\langle S_\perp^2\rangle$.
    (b) Mean excitation density $\langle n(t)\rangle$.
    (c) Logarithmic negativity $E_{\mathcal N}(t)$ for a half-chain bipartition.
    (d) Nearest-neighbor correlator $\langle N_{\rm adj}(t)\rangle$.
    Parameters from the simulation code: integrated-out cavity dynamics for the EAST constraint with periodic boundary conditions and rotating-wave approximation, $g=1$, $\kappa=30$, and $\Delta=0$. The main panels use $N=10$ with $N_{\mathrm{traj}}=200$; the inset uses $N=4,6,8,10,12,14$ with $N_{\mathrm{traj}}=80$ per point. Panel~(c) shows the mixed-state logarithmic negativity computed from the density matrix reconstructed from 200 trajectories. The superradiant burst and its quadratic peak scaling remain intact in the presence of realistic interaction tails, while the main visible effects are modest shifts of late-time observables and a mild enhancement of the generated mixed-state entanglement at intermediate tail strengths.}
    \label{fig:rydbergTails}
\end{figure}

In a realistic Rydberg implementation, the kinetic constraint is never perfectly local because the van der Waals interaction retains its $1/r^6$ tail beyond the nearest neighbor. In particular, the next-nearest-neighbor shift is inherited from the same interaction $V_{jl}=C_6/|r_j-r_l|^6$, so the ideal EAST constraint is replaced by a weak residual detuning generated by more distant excitations. The question addressed here is therefore whether these Rydberg tails qualitatively modify the constrained collective decay dynamics, or merely perturb the idealized nearest-neighbor picture. Figure~\ref{fig:rydbergTails} answers this by comparing the time evolution of the key observables for several values of the residual tail strength $V_2/\Gamma$: panel~(a) shows the collective transverse coherence $\langle S_\perp^2\rangle\equiv \langle S_x^2+S_y^2\rangle$ together with the finite-size scaling of its maximum, panel~(b) shows the mean excitation density $\langle n(t)\rangle$, panel~(c) shows the logarithmic negativity $E_{\mathcal N}(t)$, and panel~(d) shows the nearest-neighbor correlator $\langle N_{\rm adj}(t)\rangle$.

Figure~\ref{fig:rydbergTails}(a) shows that the superradiant burst survives essentially unchanged once the $1/r^6$ tails are included: the inset still exhibits the same quadratic peak scaling with system size, and the early-time dynamics is nearly indistinguishable from the ideal constrained model. The main effect appears only at late times, where stronger tails slightly suppress the stationary transverse coherence. Panel~(b) shows a similarly weak sensitivity of the one-body relaxation: the magnetization profile at early times is almost unchanged, while for stronger residual repulsion the system relaxes somewhat further at late times and reaches a slightly smaller stationary excitation density. Panel~(c) shows that the central entanglement signature of this work is fully robust. The logarithmic negativity remains large throughout the explored range of tail strengths and is even modestly enhanced at intermediate $V_2/\Gamma$, indicating that the additional interactions can increase, rather than destroy, the mixed-state entanglement generated by the decay dynamics. Finally, panel~(d) shows that the nearest-neighbor correlator is also only weakly affected: its overall relaxation profile and late-time value change quantitatively but not qualitatively. Taken together, Fig.~\ref{fig:rydbergTails} shows that realistic Rydberg tails do not alter the main physical conclusions; they leave the transient superradiance intact and preserve the entanglement-generation mechanism that underlies dark-state fragmentation.

\subsection{Dephasing}

Another experimentally relevant imperfection is dephasing noise, which can arise either locally from site-dependent fluctuations, for example due to the optical tweezers, or collectively from global phase noise of the driving laser or cavity fields. To test whether the entanglement dynamics survives under such conditions, we supplement the constrained collective decay of the main text by dephasing Lindblad terms. For individual single-site dephasing we take jump operators $L^{\rm ind}_{j,\phi}=\sqrt{\gamma_\phi}\,\sigma_j^z$, corresponding to
\begin{align}
    \mathcal L_{\phi}^{\rm ind}(\rho)=\gamma_\phi \sum_j \mathcal D[\sigma_j^z]\rho,
\end{align}
whereas for common-mode dephasing we use the main-text convention
\begin{align}
    \mathcal L_{\phi}^{\rm com}(\rho)=\gamma_\phi \mathcal D[S^z]\rho,
    \qquad S^z=\frac12\sum_j\sigma_j^z.
\end{align}
The central question is then whether the logarithmic-negativity dynamics generated by the constrained superradiant decay remains visible once these additional noise channels are included.

\begin{figure}
    \centering
    \includegraphics[width=0.8\linewidth]{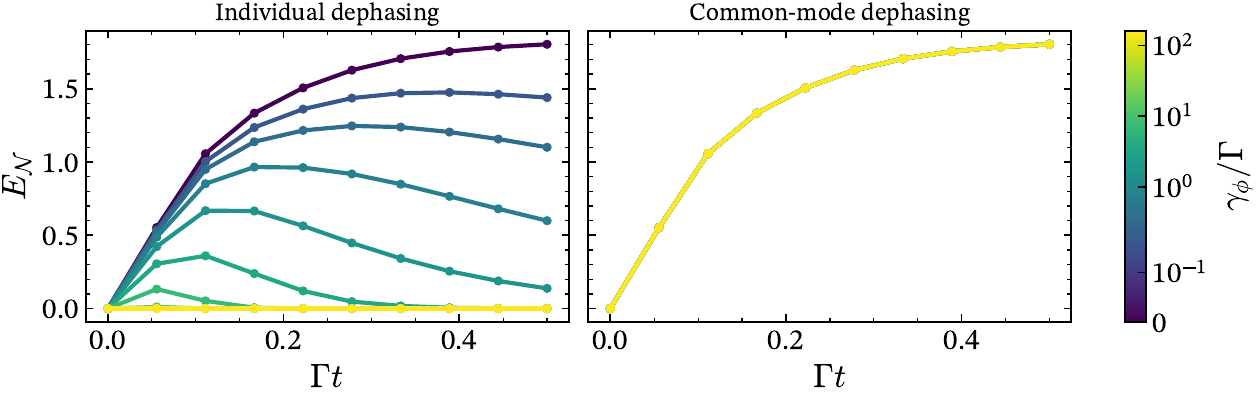}
    \caption{\textbf{Sensitivity of the entanglement dynamics to individual and common-mode dephasing.}
    Time-dependent logarithmic negativity $E_{\mathcal N}(t)$ for a half-chain bipartition in a system of size $N=8$, for a logarithmic sweep of dephasing strengths $\gamma_\phi/\Gamma$ as indicated by the color bar.
    (a) Individual single-site dephasing, with jump operators $L^{\rm ind}_{j,\phi}=\sqrt{\gamma_\phi}\,\sigma_j^z$.
    (b) Common-mode dephasing with $\mathcal L_{\phi}^{\rm com}(\rho)=\gamma_\phi\mathcal D[S^z]\rho$.
    In panel~(a), weak dephasing leaves a pronounced entanglement plateau, while stronger dephasing suppresses both the peak value and the long-time signal.
    In panel~(b), the curves collapse onto the lossless result over the full range of $\gamma_\phi/\Gamma$, showing that the entanglement-generation mechanism is essentially insensitive to common-mode phase noise.}
    \label{fig:dephasing}
\end{figure}

Figure~\ref{fig:dephasing}(a) shows that the effect of individual dephasing is controlled by the ratio $\gamma_\phi/\Gamma$. For $\gamma_\phi/\Gamma \ll 1$, the constrained superradiant dynamics still dominates: the logarithmic negativity rises rapidly and then enters a broad plateau that decays only slowly on the plotted timescale. As $\gamma_\phi/\Gamma$ is increased, however, both the peak negativity and the duration of this plateau are progressively reduced. Once $\gamma_\phi/\Gamma \gtrsim 1$, the dephasing competes with or exceeds the collective decay rate, and the entanglement is strongly suppressed already at early times. The figure therefore identifies two regimes: a superradiance-dominated window for $\gamma_\phi/\Gamma \lesssim 1$, where the characteristic entanglement dynamics survives, and a dephasing-dominated regime for $\gamma_\phi/\Gamma \gtrsim 1$, where the logarithmic negativity is rapidly washed out.

By contrast, panel~(b) shows that common-mode dephasing has essentially no visible effect on $E_{\mathcal N}(t)$, even when $\gamma_\phi/\Gamma$ is varied over several decades. Within the numerical resolution of the figure, all curves lie on top of one another and reproduce the lossless entanglement dynamics. This is an important practical result because, in cavity-based implementations, common-mode phase noise is expected to be the dominant source of dephasing. Figure~\ref{fig:dephasing} therefore shows that the dark-state-fragmentation phenomenology discussed in the main text is robust against the most relevant experimental dephasing channel, while only genuinely local phase noise poses a significant limitation.

\bibliography{bib}